\input harvmac
 \input epsf \figno=0

\def\inv{^{\raise.15ex\hbox{${\scriptscriptstyle -}$}\kern-.05em 1}}

\def\frac#1#2{{\textstyle{#1\over#2}}}

 \def\R{\relax{\rm I\kern-.18em R}}
\font\cmss=cmss10 \font\cmsss=cmss10 at 7pt
\def\Z{\relax\ifmmode\mathchoice
{\hbox{\cmss Z\kern-.4em Z}}{\hbox{\cmss Z\kern-.4em Z}}
{\lower.9pt\hbox{\cmsss Z\kern-.4em Z}}
{\lower1.2pt\hbox{\cmsss Z\kern-.4em Z}}\else{\cmss Z\kern-.4em Z}\fi}
\def\pl{Phys. Lett. B }

\lref\Iml{I.K. Kostov, Phys. Lett. B 266 (1991) 42}
\lref\ks{I. Kostov and M. Staudacher, to be published}
\lref\Icar{I.Kostov, ``Strings embedded in Dynkin diagrams'',  Lecture 
given at the Cargese meeting, Saclay preprint
SPhT/90-133}
\lref\Mig{A.A. Migdal, Phys. Rep. 102 (1983) 199}
\lref\bkz{E. Br\'ezin, V. Kazakov, and Al. B. Zamolodchikov,
\Nucl. Phys. B338 (1990) 673}
\lref\Iade{I. Kostov, Nucl. Phys. B 326 (1989)583}
\lref\bkkm{D. Boulatov, V. Kazakov, I. Kostov, and A.A. Migdal, Nucl. Phys. B
275 [FS 17] (1986) 641}
\lref\Inonr{I. Kostov, \pl 266 (1991) 317.}
\lref\bpz{A. Belavin, A. Polyakov, and A. Zamolodchikov, Nucl. Phys.
B 241 (1984) 333}
\lref\df{V.Dotsenko and V. Fateev, Nucl. Phys. B 240 (1984) 312}
\lref\Imult{I. Kostov, 
\pl 266 (1991) 42.}
\lref\ajm{J. Ambjorn, J. Jurkiewicz and Yu. Makeenko, \pl 251 (1990)517}
\lref\mat{V. Kazakov, Phys. Lett. 150B (1985) 282;
F. David, Nucl. Phys. B 257 (1985) 45;
V. Kazakov, I. Kostov and A.A. Migdal, Phys. Lett. 157B (1985),295;
J. Ambjorn, B. Durhuus, and J. Fr\"ohlich, Nucl. Phys. B 257 (1985) 433}
\lref\brkz{E. Br\'ezin and V. Kazakov, Phys. Lett. 236B (1990) 144.}
\lref\dglsh{M. Douglas and S. Shenker, Nucl. Phys. B 335 (1990) 635.}
\lref\abf{G. Andrews, R. Baxter , and P. Forrester, J. Stat. Phys. 35
 (1984) 35}
\lref\bax{R. Baxter, Exactly Solvable Models in Statistical Mechanics
(Academic Press, London,1982)}
\lref\grmg{D. Gross and A. Migdal, Phys. Rev. Lett. 64 (1990) 127.}
\lref\mike{M. Douglas, Phys. Lett. 238B (1990) 176.}
\lref\pol{A. Polyakov, Phys. Lett. 103 B (1981) 207, 211.}
\lref\kpz{V. Knizhnik, A. Polyakov and A. Zamolodchikov, Mod. Phys. Lett.
A3 (1988) 819.}
\lref\ddk{F. David, Mod. Phys. Lett. A3 (1988) 1651; J. Distler and
H. Kawai, Nucl. Phys. B 321 (1989) 509}
\lref\bk{M. Bershadski, I. Klebanov, Nucl. Phys. B 360 (1991) 559 }
\lref\mss{G.Moore, N.Seiberg and M. Staudacher, Nucl. Phys. B 362 (1991)665}
\lref\polci{J. Polchinski, Nucl. Phys. B346 (1990) 253 }
\lref\stau{M. Staudacher,Nucl. Phys. B 336 (1990) 349}
\lref\bdks{E. Br\'ezin,
M. Douglas, V. Kazakov, S. Shenker, Rutgers preprint RU-89-47 (1989)} 
\lref\gm{D. Gross and A. Migdal, Nucl. Phys. B340 (1990) 333}
\lref\David{F. David, Mod. Phys. Lett. A, No.13 (1990) 1019} 
\lref\bax{R. J. Baxter, Exactly solved Models in Statistical Mechanics
(Academic Press, 1982) } 
\def\CC {{\cal C}}
\def\CV {{\cal V}}
\def\l {\ell }
\def\L {\Lambda }

\def\tr {\rm tr}
\def\gst {\gamma _{\rm str }} 
\def\p {\partial}
\def\CS {{\cal S}}

\def\CA {{\cal A}}

\def\CR {{\cal R}}
\def\CP {{\cal P}}
\def\CD {{\cal D}}

\def\c {\circ }
 \rightline{hep-th/9112059}
\Title {} {Strings with  Discrete Target  Space}

\centerline{Ivan K. Kostov \footnote{$^\ast $}{on
 leave of absence from the Institute for Nuclear
 Research and Nuclear Energy, Boulevard Trakia 72, BG-1784 Sofia,Bulgaria}}

\centerline{{\sl Service de Physique Th\'eorique}
\footnote{$ ^\dagger$}{
Laboratoire de la Direction des Sciences de la Mati\`ere du
Commissariat \`a l'Energie Atomique.} {\sl de Saclay, }
{\sl CEN-Saclay, F-91191  Gif-sur-Yvette , France}}

\vskip .3in
\baselineskip12pt{\ninepoint 
We investigate the field theory of strings having as a target space
an arbitrary  discrete
 one-dimensional manifold. The existence of the continuum limit
is guaranteed if the target space is a Dynkin diagram of a simply laced
Lie algebra or its affine extension. In this case the  theory
 can be mapped
onto the theory of strings embedded in the 
 infinite discrete line $\Z$ which is the target space of the
SOS model. On the regular lattice this mapping is known as Coulomb gas
picture.  Introducing a quantum string field $\Psi _{x}(\l )$ depending
on the position $x$ and the length $\l$ of the closed string,
 we give a formal definition of the string field theory
in terms of a functional integral.    The classical string background
 is  found as a solution of the saddle-point equation which is equivalent
to the loop equation we have previously considered \Iade . 
The continuum limit exists in the vicinity of the singular points of this
equation. We show that for given target space
 there are many ways to achieve the continuum limit; they are
 related to the
multicritical points of the ensemble of surfaces without embedding.  
 Once the classical background is known, the  amplitudes involving
propagation of strings
 can be evaluated  by perturbative expansion around the saddle point
of the functional integral. For example, the partition function
of the noninteracting closed string (toroidal world sheet) is
 the contribution of the
gaussian fluctuations of the string field.
The vertices in the corresponding Feynman diagram technique
 are  constructed as  the loop amplitudes in a random matrix
model with suitably chosen potential. 
}
\vskip 1cm
\leftline{ {Nuclear Physics}  B 376 (1992) 539-598}
 \rightline{SPhT/91-142}

\Date{9/91}

\baselineskip=12pt

\vskip .3in

\newsec{Introduction}

The recent progress in the theory of non-critical strings
 has been prepared by the conjecture that the 
functional integral of 2d gravity can be discretized as a sum over
planar graphs \mat     \ . Such a discretization allowed to apply
powerful methods of calculation borrowed from the theory of random
matrices \ref\bipz{E. Br\'ezin, C. Itzykson, G. Parisi and J.-B.
Zuber, Comm. Math. Phys. 59 (1978) 35; D. Bessis, C. Itzykson and J.-B. Zuber,
 Adv. Appl. Math. 1 (1980) 109; C. Itzykson and J.-B. Zuber, J. Math. Phys. 21
(1980) 411}. Some of the corresponding matrix models can be solved
exactly after being reduced to  a problem of non-interacting fermions
in a common potential. This fermionic representation proved to be very
helpful in practical calculations but its connection with the original
model seems quite formal.
It is not clear, for example,  whether the nonperturbative phenomena  
in the fermion problem can be given an interpretation in terms of strings.

It is therefore desirable to have a formalism in which the string excitations
appear explicitely. An example of such a formalism  applied to the 
string propagating in the one-dimensional  space $\R$
is the collective field
approach of Das and Jevicki  \ref\dj{S. Das and A. Jewicki, Mod. Phys.
Lett. A5 (1990) 1639; K. Demeterfi, A. Jevicki and J. Rodrigues, Brown 
Univ. preprint, BROWN-HET-795, February 1991 }. 

In this paper we develop a similar formalism for  string theories
 with discrete target spaces. 
Bosonic string with discrete target spaces are interesting mainly because of 
their interpretation as theories of two-dimensional gravity coupled
to matter. The (discrete) degrees of freedom of the matter field are
the points of the target space. Because of the enormous symmetry of the
problem, it is much easier to investigate the critical behaviour
of statistical systems on a fluctuating surface than on the plane. 
The exact results obtained for a fluctuating surface can be translated
according to \kpz , \ddk  \ to the case of a frozen geometry of the 
world sheet.

 The  string models we are investigating are
a generalization of 
the $ ADE $  strings constructed in ref. \Iade .
They are closely related to  the  SOS and RSOS solvable statistical
models on a regular planar lattice \ref\bax{R. Baxter, Exactly Solved
 Models in Statistical Mechanics, Academic Press, 1982}. 
    The target spaces
of these models are
one-dimensional discrete manifolds representing Dynkin diagrams of 
simply laced Lie algebras or their affine extensions.
All such target spaces can be mapped onto a master target space
which is the infinite discretized line $\Z$.   

Since in a one-dimensional embedding space there is no room for 
transverse excitations, the states of the closed
 string is completely determined
by the two global modes - the center of masses $x$ and its length $L$.
The string field $\Psi (L)$ is an operator creating on the world sheet
 a boundary of length $L$  and position $x$ in  the target space. The
path integral for the correlation functions of the
 string field involve embeddings of the world sheet such that
each connected component of its boundary is mapped into a single point
of the target space. 

The string models related to the SOS model are remarkable with the
possibility to reduce the string path integral
 to a simpler one involving only the global 
modes $x$ and $L$, even {\it before} going to the continuum limit.
As a result one obtains an effective two-dimensional QFT for the
field $\Psi _{x}(L)$. 

The reduction of the string path integral stems from the loop, or polygon
representation of the SOS-type statistical systems \bax\ which has
been generalized to the case of irregular lattices in \Iade .
 A map of the 
world sheet of the string in the target space can be represented by
a collection of nonintersecting loops (domain walls) separating the
domains of constant coordinate $x$. We assume that the discontinuity 
across a domain wall is  $\pm 1$. Each loop configuration is characterized by 
its topology, the lengths of the loops, and the coordinates of the connected
domains. It can be imagined as a ``stroboscopic picture'' of a
the evolution of one or several closed 
strings, the time direction on the world sheet 
being at all points orthogonal to the domain walls.    
 The evolution of the string is decomposed into elementary processes
 represented 
by the domain walls and connected domains.  
The domain walls  describe  elementary steps in the $X$-space without changing
the world-sheet geometry. On the other hand, the connected domains
describe topology-changing processes
without propagation in the embedding space. This nice factorization 
renders the model exactly solvable. 

A loop configuration is represented by a graph with lines corresponding
to domain walls and vertices corresponding to the connected domains on the
world sheet.   
There are vertices of all orders including tadpoles describing
processes of  creation or decay of a closed-string state. Each vertex
is characterized also by the number $H$ of handles of the
connected domain on the world sheet. Vertices with $H>0$ describe
processes involving creation and annihilation of $H$ virtual closed strings
at the same point $x$ of the embedding space. 

The string path integral can be written as a sum over all loop configurations, 
and the entropy of a loop configuration factorizes 
to a product of weights associated with the domains. The weight of each
domain is proportional to the entropy of a {\it nonembedded} random
surface with the corresponding topology. The latter can be calculated,
for example, by using the matrix model formalism.  
 In this  way the
the graphs representing the loop configurations can be interpreted as
Feynman diagrams for 
 an ordinary field theory in the space of global modes  $(x,L)$
 and the string path
integral is reduced to a functional integral for the field $\Psi _{x}(L)$.

The first (and in a sense the most difficult) problem to solve is to find
the classical 
string background  $\Psi _{x}^{cl}(L)$ 
which is the solution of the classical equation of motion in 
the effective field theory.  Then the propagation of the noninteracting closed
string is determined by the gaussian fluctuations of the string
field around the classical solution and the vertices are given by the 
higher order terms.
The accomplishment of this program will be the subject of this paper.

We  are going to present in details the results announced in three
short publications \Icar , \Inonr , and  \Imult , as well as a few new
results as the microscopic construction of the dilute $ADE$ models
 and a description of
their multicritical points \ks .  

The paper is organized as follows. 
\smallskip
In section 2 we define the path integral for a string theory with discrete
target space and  establish its equivalence with an ordinary field theory
for the string field $\Psi _{x}(L)$.
The measure in the space of world sheet geometries will be discretized
by planar graphs. Therefore we have first to define  the statistical
systems related to the SOS model (known also as
 interaction-round-a-face, or IRF
models \bax )  on an arbitrary irregular lattice $\CS$ .
In sect. 2.1 we present a generalized version of the construction of these 
models worked out in \Iade . The present version allows two nearest neighbour 
sites $s$ and $s'$ of $\CS$ to have coordinates (heights) $x$ and $x'$
which either  or coincide or  are  nearest neighbours in the $X$ space.
In this way we are able to achieve both the dense and dilute phases of
the IRF models.  The important  special case $X=\Z$ is
discussed in sect. 2.2. where we  consider a nonunitary version of the SOS
model with complex Boltzmann weights. This model renormalizes at
large distances onto a gaussian field with a linear term coupled to the local
curvature of the discretized 
world sheet.  The mapping of the IRF models onto this
SOS model known as Coulomb gas picture, 
can be constructed in the same way as in the case of a lattice
without curvature \bax , \ref\fsz{P. di Francesco, H. Saleur, and
J. B. Zuber, Journ. Stat. Phys., 49 (1987) 57}, \ref\fn{O.Foda
 and B. Nienhuis, Nucl Phys.B 324 (1989) 643}. The  Coulomb gas picture
is based on the loop expansion which is explained in sect. 2.3.
The partition function of the statistical system with target space $X$ 
equals the sum over the embeddings of a world sheet with fixed geometry.
In section 2.4 we define the string path integral
 as a  sum over the world sheet geometries and review its description 
by means of a two-component gaussian field \pol  \kpz \ddk . The normalization
of the electric and magnetic charges is fixed by the  correspondence
with the SOS model. In section 2.5 we establish  the equivalence 
between the string 
path integral  and the Feynman diagram  series for the field $\Psi _{x}(L)$
describing the dynamics of the global modes of the string.
By means of  a shift $\Psi _{x}(L)  \to
\Psi _{x}(L)+\Psi^{cl} _{x}(L)$  we define an improved diagram technique
containing no tadpoles.  The classical string 
background $\Psi ^{cl}_{x}(L)$ is determined by the saddle-point equation.
In sect. 2.6 we show that this equation is equivalent to the loop 
equation considered in \Iade .
 \smallskip
In section 3 we study the continuum limit of the string field theory.
For this purpose we introduce a cutoff $a$ with dimension of length and
replace the dimensionless length $L$ with a  renormalized length
$\l = a L$. The continuum limit can be achieved if the bare parameters of
the string are tuned to a critical point.
The critical points can be found as the singularities of the loop equation
for the classical string background. The latter allows an interpretation
as the partition function of a gas of nonintersecting loops on a 
fluctuating surface with the topology of a disc. 
In the vicinity of a critical point the volume of the world sheet 
always diverges and the different critical regimes
  are distinguished by
the behaviour of the loops. The latter is gouverned by a parameter 
(energy) coupled to the total length of the loops. If the
energy  exceeds the entropy, then
 the loops remain  small, and  the possible
critical regimes are the $m$-critical points of nonimbedded surfaces
(sect. 3.1). This is the phase of noncritical loops.  
In the opposite case, considered in sect. 3.2, 
 the energy of the loops is not sufficient
to compensate their entropy, and the loops
  form a dense critical  phase filling the world sheet
(sect. 3.2). Finally, the dilute phase of the loop gas (sect. 3.3) is achieved
when the energy of the loops is tuned to its critical value.
In the dilute phase the loops are still critical but they cover only a
small fraction of the points of the lattice.
In this phase the critical behaviour is sensitive to the choice of the
integration measure in the space of world sheet geometries. By tuning
the Boltzmann weights of the planar graphs we can achieve an infinite 
sequence of multicritical points of the dilute phase.  
The continuum limit of the loop equation 
 (sect. 3.4) is universal ; the different critical regimes
are classified by  the possible  asymptotics
 of the solution at small lengths. 

\smallskip

Once the string background is known, it is not difficult to
find explicit expressions for the Feynman rules in the continuum
limit.  The vertices in the improved diagram technique (i.e.,
dressed by tadpoles) can be calculated as the loop correlators in
a special matrix model constructed in sect. 4.1. The physical meaning 
of the coupling constant of this matrix model is discussed in sect.
4.2. It is coupled to a local operator which has negative dimension 
if the theory is not unitary.
Its dimension is related to the fractal dimension of the connected
domains on the world sheet.
In sect.4.3 we give explicit expressions for the planar ($H=0$) vertices
in the continuum limit. Since the nonplanar loop amplitudes in the
one-matrix model are not yet known, we have only fixed the general structure 
of the nonplanar vertices. The expression for a vertex with $H$ handles
and $n$ legs contains $3H+n-3$ unknown coefficients.
The Feynman rules simplify in the momentum space $(p,E)$ where
the propagator of the string field diagonalizes. The spectrum of on-shells 
states
discussed in sect. 4.4  is given by the light cone in
 the $(p,E)$ space. The latter 
has the geometry if a half-infinite cylinder since the discreteness of
the $X$ space leads to periodicity in the $p$-direction.
  Finally, in sect. 4.5 we check that the partition function of the 
noninteracting string (closed surfaces with the topology of a torus)
is obtained by integrating over the gaussian fluctuations around the saddle
point.

\newsec{ Statistical systems on planar graphs and
 kinematics of strings with discrete target space}
\subsec{The IRF height models on an arbitrary
irregular lattice}

   It is well known \ref\ciz{A. Cappelli, C. Itzykson and J. B. Zuber,
Comm. Math. Phys. 113 (1987) 113} that the  two-dimensional rational
 conformal-invariant QFT with central charge $C < 1$ are classified by the
simply laced Lie algebras (i.e., these of the classical series $ A_{n},
D_{n}, E_{6},E_{7},E_{8}$). 
 Each of these theories can be constructed microscopically 
as a lattice statistical model whose local degrees of freedom are labeled by
the points of  the Dynkin diagram of the corresponding Lie algebra 
\ref\pas{V.
Pasquier, Nucl. Phys. B285 (1987) 162; J. Phys. 35 (1987) 5707}.
The statistical models associated with dynkin diagrams of $A,D,E$
type, or shortly 
$ADE$ models  represent a natural generalization
 the RSOS face models considered by Andrews,
Baxter and Forrester \abf . The local degrees of freedom are attached to the
sites of the lattice and interact through ``interactions-round-a-face''
(IRF) around each plaquette.  
 
 Similarly, the
extended  $\hat A \hat D \hat E$  Dynkin
 diagrams describe conformal invariant QFT with $C=1$ and discrete spectrum
of conformal dimensions. In fact, Pasquier's construction \pas\ 
 can be applied 
to any one-dimensional discrete manifold $X$ , i.e.,a set of points $x$ and
links $\langle x x' \rangle $   with the structure
of a one-dimensional simplicial complex.  The requirement that $X$ is an
(extended) Dynkin diagram of ADE type guarantees the existence of a scaling
limit. Another important target spase is the infinite discretized line $\Z$.
This is the target spase of the model known as SOS (solid-on-solid) model.

Before presenting the definition of the IRF model with target space $X$
we are going to describe the space of excitations of this model.

The one-dimensional discrete manifold $X$ is defined by the set of its
points x and the adjacency
matrix $C_{xx'}, \ x,x' \in X$
\eqn\cnctmt{C_{xx'}=[{\rm the\  number\  of\ 
 links\  connecting\  }x\  {\rm and }\ x']}
We assume that the target space $X$ is represented by a nonoriented 
graph which implies symmetric adjacency matrix.
 
The Hilbert space of states
 of the $X$-field  in a fixed ``time slice'' consists of all closed
paths in $X$ with given length.
  Therefore in order to identify the ground state and the
excited states we have to solve the problem of random motion in
$X$. The propagation kernel for a random walk on $X$ consisting of $n$
steps is just the $n$-th power of the adjacency matrix $C$.
Introducing the eigenvectors $V^{x}_{(p)}$
\eqn\eIi{\sum_{x'} C_{x x'} V^{x'}_{(p)}= \beta_p
V_{(p)}^{x'} }
we can write the kernel $K_{xx'}=(C^{n})_{xx'}$ as a sum of projectors on the
eigenstates
\eqn\eIii{(K^{n})_{xx'}=\sum_{p}(\beta_p)^{n}V^{x}_{(p)}V^{x'}_{(p)}}

If $X$ is an $ADE$ Dynkin diagram (Fig.1),
  then the eigenvalues of $C$ have the form
\eqn\eIiii{\beta_{p}=2\cos (\pi p),  \enspace p=m/h}
where $h$ is the Coxeter number and the integer 
 $m$ is one of the Coxeter exponents of
the Dynkin diagram \foot{The inverse statement reads: if
 all the eigenvalues of the adjacency matrix are less
 than 2, then the graph $X$ is an either an 
$ADE$ dynkin diagram or a quotent $A_{2n}/Z_{2}$ \ref\goodm{F.M. Goodman,
P. de la Harpe and V.F.R. Jones, Coxeter - Dynkin diagrams and towers of
algebras, vol. 14, Mathematical Research Institute Publications, Springer
Verlag 1989}}  . The eigenvectors $V^{x}_{(p)}$ define the Fourier
transform from coordinates $x$ to the discrete momenta $p=m/h$. 
The ground state corresponds to the maximal eigenvalue 
of the adjacency matrix $C$
\eqn\manev{\beta \equiv \beta _{p_{0}} =2\cos (\pi p_{0}), \ \ \  p_{0}=1/h
\ \ \ \ \ \ \ \ (A,D,E) }

  \epsfxsize=390pt
 \vskip 20pt
 \hskip -20pt
 \epsfbox{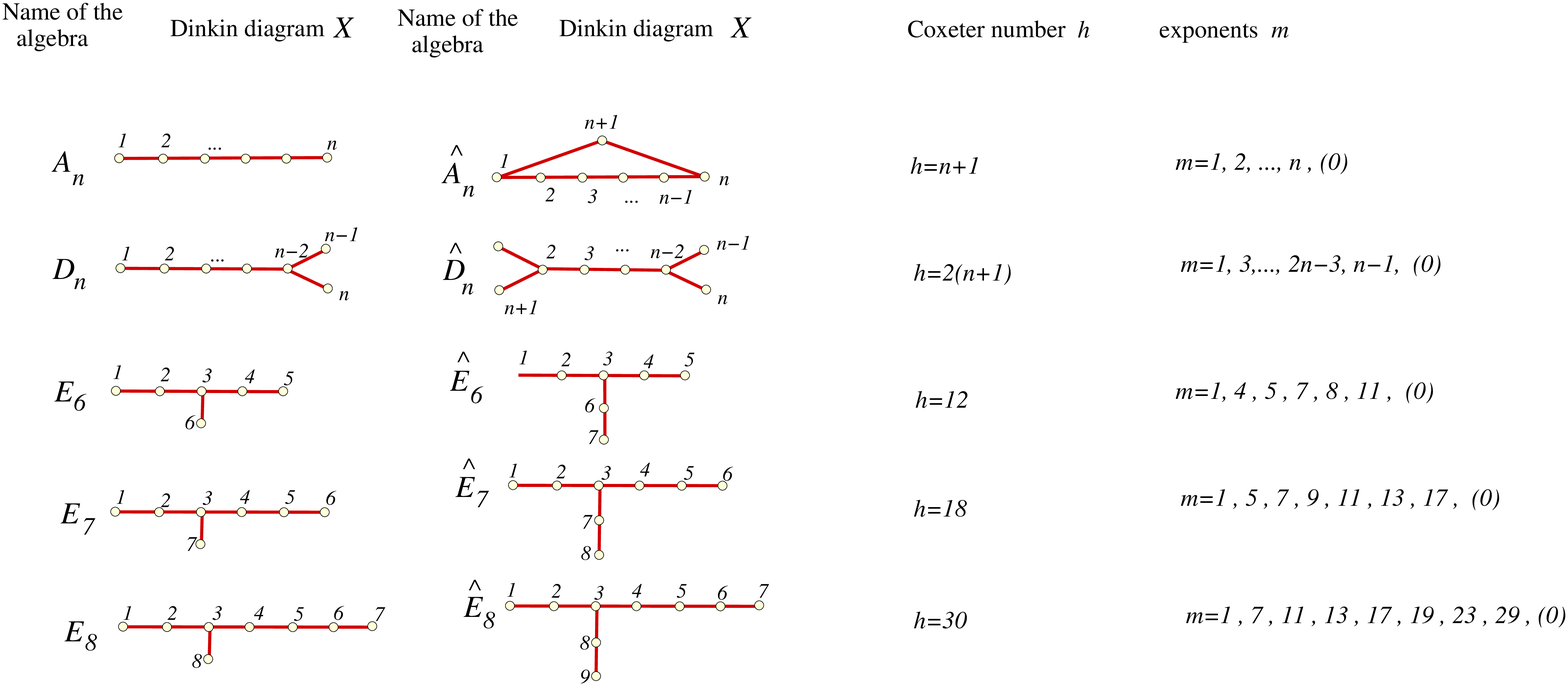}
 \vskip 5pt
 
 \centerline{\ninepoint Fig.1 :   
 Dynkin diagrams   and their Coxeter exponents }
  
  \vskip  30pt

If  $X$ is an  extended  $\hat A \hat D \hat E$  Dynkin diagram, then 
the spectrum of Coxeter exponents includes $m=0$ and the maximal eigenvalue 
of $C$ is
\eqn\bbet{\beta =2, \ \ \  p_{0}=0 \ \ \ \ \ \ \ 
\ \ \ \ \ \ \ \  \ (\hat A, \hat D, \hat E ) }
In order to
simplify the notations we shall always denote the ground state
wavefunction by
\eqn\sxvx{S_x = V^{x}_{(p_{\c})}}
Since the entities of the adjacency matrix are nonnegative, by the
Peron-Frobenius theorem $S_{x}\ge 0$ for all $x \in X$.

The propagation kernel \eIii\ becomes in the limit of 
large proper time $n$  a single
projection operator to the ground state $S_{x}$
\eqn\proj{(K^{n})_{xx'} \sim S_{x} \ \ ( \beta ) ^{n}\ \  S_{x'}}
It is convenient to define the wavefunctions of the excited states
as
\eqn\eIv{\chi ^{x}_{(p)}=V^x_{(p)}/S_x}
They satisfy a closed algebra 
\eqn\eIvi{\chi ^x_{(p)} \chi ^x_{(p')} =\sum _{p''}C_{pp'p''} \chi
^x_{p''}}
 and  orthogonality conditions of the form
\eqn\eIvii{\sum _{x} S_{x}^{2} \chi _{(p)}^{x} \chi _{(p')}
 ^{x}= \delta _{pp'} , \ \ \sum _{p} \chi _{(p)}^{x} \chi _{(p)}^{x'}
=\delta _{x,x'} (S_{x}S_{x'})^{-1}}

Let us give two simplest examples.
\smallskip
 
{\it i}) The Dynkin diagram of the algebra $A_{h-1}$  is 
the chain of $h-1$ points $x=1,2,...,h-1$. This is the target space of the
RSOS models \abf . The eigenvectors of $C$ are 
\eqn\eIiv{V^{x}_{(p)}=\sqrt {{2 \over h}}\ \sin (\pi px), \  p=1/h,
2/h,...,(h-1)/h;\ \ \ \ \ S_{x}=\sqrt {{2 \over h}}\  \sin (\pi x/h)}
In this case the ground state corresponds to $p_{0}=1/h$.
  
 \smallskip
 
{\it ii}) The extended Dynkin diagram  corresponding to the
affine Kac-Moody algebra  $\hat A_{2h-1}$ is the
ring $ \Z_{2h}$  made of $2h$ points. It 
can be represented by the set of integers 
modulo  $2h$. Then the connectivity matrix $ C_{xx'}$ reads
\eqn\adjm{ C_{xx'} = \delta ^{(2h)}_{x,x'+1} + \delta ^{(2h)}_{x,x'-1}}
where  $\delta ^{(2h)}$ is the Kronecker symbol modulo $2h$. The eigenvectors
of the matrix $\hat C$ are 
\eqn\eigz{V_{(p)}^{x} = ( 2h)^{-1/2} \exp(i \pi px), p=0, \pm {1 \over h},...,
\pm {h-1 \over h},1. }
The target space is translational invariant and the spectrum of momenta 
contains the point $p=0$. The ground state is $S^{x}=V_{(0)}^{x}=1/\sqrt {2h}$.
The vertex operators (order parameters) 
\eqn\ordz{ \chi _{(p)}^{x} = e^{i \pi px}}
satisfy a closed algebra with  fusion rules representing the momentum 
conservation modulo the period 2  of the momentum space
\eqn\fzrz{C_{pp'p''} = \delta ^{(2)}( p+p'+p'')}

The microscopic definition of the  IRF models on a general irregular lattice
was  presented in \Iade\  as a direct generalization of Pasquier's 
construction \pas   , the only new point being the explicit dependence of the 
Boltzmann weights on the scalar curvature. The effects of the curvature 
have been also studied in  \fn . 
  Below we  present a more general construction
including both the dense and the dilute versions of the IRF models.

Let $\CS$ be a two-dimensional discrete manifold such that all its faces
are squares (we shall call them also plaquettes). It can be constructed 
from an arbitrary planar graph and its dual one by taking the points of both 
graphs and adding bonds connecting  all pairs of points
which serve as extremities of mutually dual lines. A small section of
such lattice is shown in Fig.2

Each configuration of the $x$-field on $\CS$ defines a map $\CS \to X$.
 Sometimes the local values of the
$x$-field are called heights. 
The map is continuous in the sense that nearest neighbours in $X$ are 
images of nearest neighbours in $\CS$.
 In the original papers a stronger 
condition has been imposed, namely, that the links  $\langle ss' \rangle $
of $\CS$ are mapped into links $\langle xx' \rangle $ of $X$. Here we 
relax this condition by allowing two nearest neighbours in $\CS$ to be
mapped into the same point $x \in X$.

  \epsfxsize=150pt
 \vskip 20pt
 \hskip 80pt
 \epsfbox{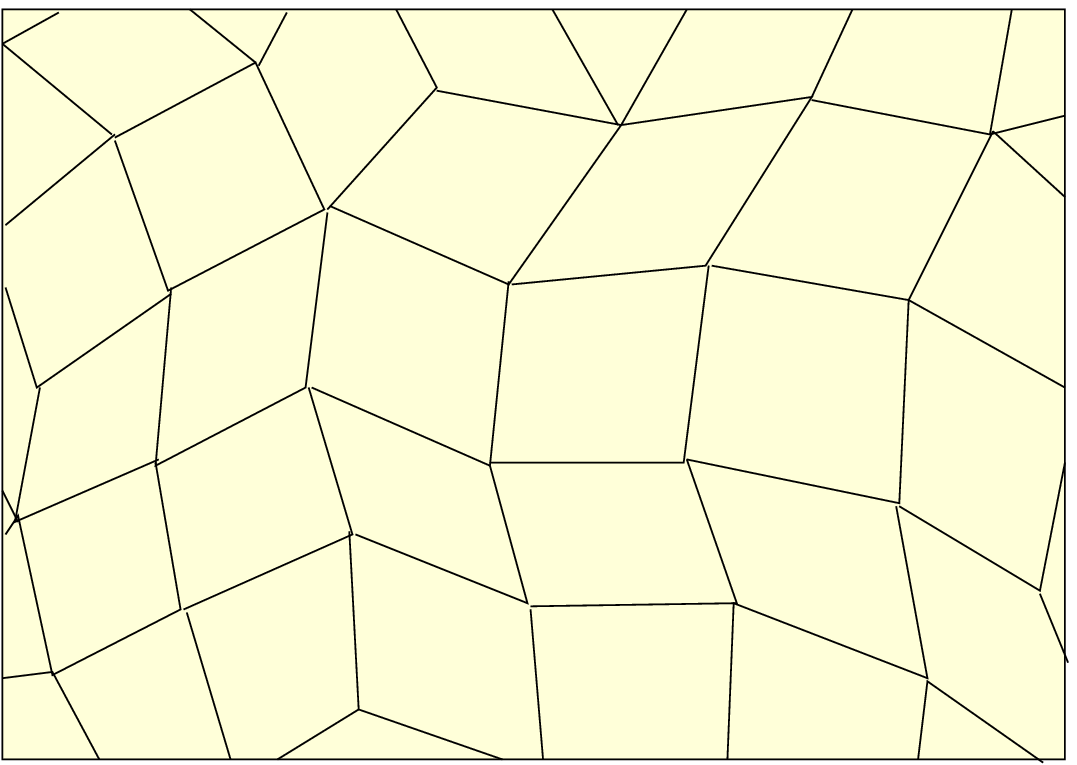  }
 \vskip 5pt
 
 \centerline{\ninepoint Fig. 2:   A section of the discretized world sheet      }
  
  \vskip  10pt

The statistical weight of each field configuration is a  product of factors
associated with the plaquettes and sites of the lattice $\CS$. 
We assume that the weights of the plaquettes are symmetric under cyclic
permutations.This property will be of crucial importance for the generalization
of the models to the case of  an irregular lattices. 
The weight of the plaquette ($ _{s_{1} s_{3}}
  ^{s_{2} s_{4}}$) is defined by
\eqn\wghts{
 \epsfxsize=  250pt
\epsfbox{   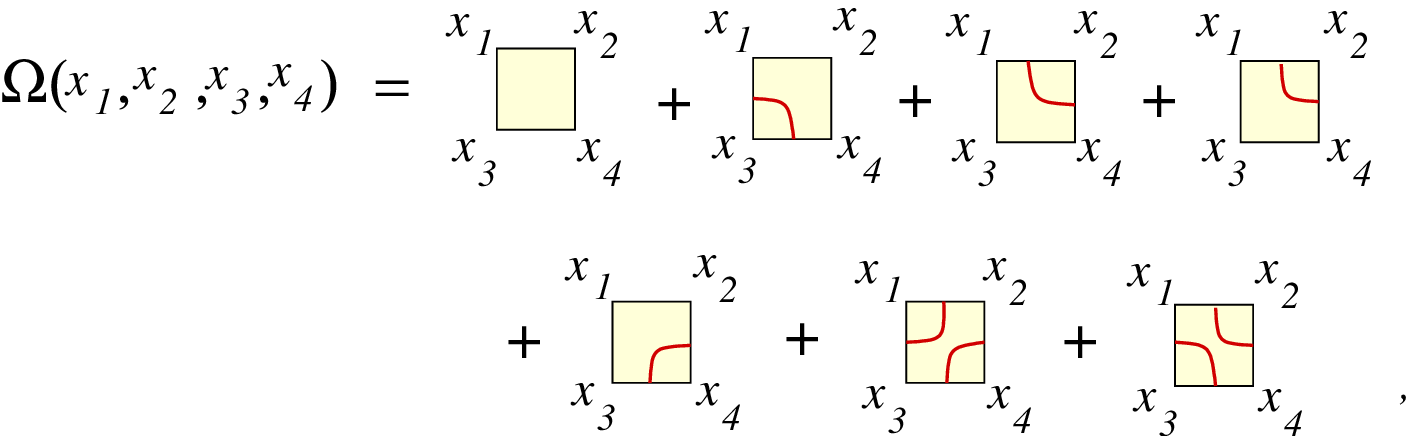 }
 }

%
with
\eqn\weightsp{\eqalign{
 \epsfxsize=  30pt
  \epsfbox{   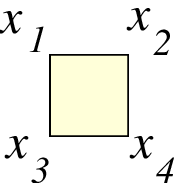 } & \qquad = \delta _{x_{1}x_{2}}\delta _{x_{2}x_{3}}
\delta _{x_{3}x_{4}}\delta _{x_{4}x_{1}}
 \cr
\epsfxsize=  30pt
  \epsfbox{   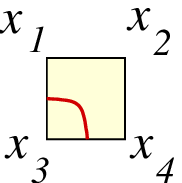 }
 & \qquad ={1 \over T} C_{x_{1}x_{3}} C_{x_{3}x_{4}} 
   \delta _{x_{4}x_{2}}
  \delta _{x_{2}x_{1}}
 \root 4 \of {{S_{x_{1}} \over S_{x_{3}}}}  \cr
 \epsfxsize=  30pt
  \epsfbox{   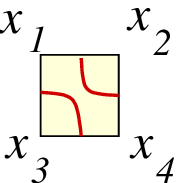 }
 & \qquad 
   = {1 \over  T^{2}}C_{x_{1}x_{3}} C_{x_{3}x_{4}} 
C_{x_{4}x_{1}} C_{x_{1}x_{2}} \ \delta _{x_{2}x_{4}} 
\root 4 \of {{S_{x_{1}} S_{x_{3}} \over
 S_{x_{2}} S_{x_{4}}}} \cr}}
where $T$ is the temperature of 
the statistical  system. 

The factors 
associated with the sites $s \in \CS$ reflect the response of the $x$-field to
the local curvature $\hat R_{s}$  which is
 concentrated at the sites. On an irregular lattice the local
Gaussian curvature at the point $s$ is defined as twice the deficit angle
\eqn\cur{\hat R_{s} = \pi (4-q_{s})}
where $q_{s}$ is the
coordination number at the site $s$ (= the number of plaquettes sharing
this site). We use the standard normalization such that 
\eqn\gbe{\sum _{s \in \CS} \hat R_{s} = 4 \pi \chi}
where $\chi = 2-2H$ is the Euler characteristic of the surface $\CS$.
The weight associated with a site $s$ is
\eqn\wsite{ \Omega (x) = (S_{x})^{\hat R_{s}/4\pi}}
The factors \wsite are trivial on a regular square lattice where all 
coordination numbers are equal to 4.

Collecting all factors, we define the partition function of the model with
target space $X$ on the lattice $\CS$ as the sum over all maps
$\CS \rightarrow X$
\eqn\partfade{F[\CS]= \sum _{ \CS \to X}\prod_{ s} \Omega (x(s))
 \prod _{ ( ^ {s_{2} s_{3}} _{s_{1}s_{4}} )}
  \Omega (x(s_{1}),x(s_{2}),x(s_{3}),x(s_{4}))}

In the limit $T \to 0 $  the r.h.s. of 
\partfade \ is  the partition function of
the critical $ADE$ models 
 considered in \Iade \ . 
In the high temperature  limit $\ T \to \infty\  $ the $x$-field
 freezes to a constant and there
are no long range correlations. At some critical temperature
 $T_{c}$ there 
should be a phase transition  between the low-temperature (critical)
and the high-temperature (noncritical) phase. At the point $T_{c}$
 a new critical regime can be achieved. 
We call this phase, by reasons which will become clear later, 
{\it dilute phase}. The critical low-temperature
phase  will be refered also as the {\it dense phase } of the IRF models.

\subsec{ The nonrestricted SOS model with a false vacuum}

It is known  \bpz\ , \df\ , that the dynamics of
 the two-dimensional conformal theories
 can be described in terms of a Gaussian massless field
 with a linear
term coupled to the local curvature.  Let  $\xi _{a},\  a=1,2$, be the local 
coordinates on a surface $\CS$ with  metric $\hat g_{ab}(\xi );a=1,2$ and local
curvature $\hat R (\xi )$. Then the gaussian 
field $x(\xi )$ is defined by the action
\eqn\gauact{\CA [x;\hat g ] = {1 \over 4 } \int d^{2}\xi
 \sqrt{{\rm det}\hat g} \ 
 [\pi g \  \hat g^{ab}\p _{a}x(\xi ) \p_{b} x (\xi)  +i \alpha _{0}
x (\xi)\hat R(\xi )]  }
where $g$ is the coupling constant and $\alpha _{0}$
is the  background electric charge.

One can map all conformal field theories with conformal anomaly $C \le 1$ 
onto this gaussian field in the sense that the partition and correlation 
functions can be interpreted in terms of distributions of electric and 
magnetic charges in the gaussian theory. This mapping is known as 
Coulomb gas picture 
  \ref\kkn{L.Kadanoff, J. Phys. A 11 (1978)1399;
H. Knops, Ann. Phys. 128 (1981) 448; M. den Nijs, Phys. Rev. B27 (1983)1674},
\ref\bn{B. Nienhuis, in Phase Transitions and Critical Phenomena,
 Vol. 11, C.C. Domb and J.L. Lebowitz, eds.) Ch. 1 (Academic Press, 1987)},
\ref\df{V.Dotsenko and V. Fateev, Nucl. Phys. B 240 (1984)312; 251 (1985)691},
\ref\gn{J.-L. Gervais and A. Neveu, Nucl. Phys. B 199 (1982) 50}.
The electric charge $e$ is carried by the vertex operator $V_{\alpha}(\xi)=
\exp (i \pi \alpha x(\xi))$ and the  charge $\mu$ at the point $\xi$
creates a discontinuity $2\mu$ along a line starting at the point $\xi$ (in
fact, the discontinuity can be distributed among several lines forming a
star with center at $\xi$). 

The (++)-component of the energy-momentum
 tensor is
%
\eqn\emt{T(\xi )= -g\pi ^{2} \p _{\xi}x \p _{\xi}x +i\pi \alpha _{0}\p _{\xi}
^{2} x(\xi )}
where we have introduced the complex variable $\xi _{\pm}= \xi_{1}\pm 
i\xi _{2}$.
From the o.p.e. \  \bpz , \df
\eqn\ope{T(\xi )T(\xi ')={1 \over 2}{1 \over (\xi -\xi ')^{4}}
[1-6 { \alpha _{0}^{2} \over g}] \ +{2 \over (\xi -\xi ')^{2}} T(\xi _{2})+...}
\eqn\cdgf{T(\xi )V_{\alpha}(\xi ') = {\alpha (2\alpha _{0} -\alpha )
\over 4g} {1 \over (\xi -\xi ')^{2}} V_{\alpha}(\xi ') +...}
one reads that the conformal anomaly $C$ of the field \gauact\  and the
conformal dimensions of the vertex operators $V_{\alpha }$ are
\eqn\conan{c=1-6{\alpha _{0}^{2} \over g}; \ \ \ 
\Delta _{\alpha} = {\alpha (\alpha - 2 \alpha _{0}) \over 4g}}
The operators $V_{\alpha} $ and $V_{2\alpha _{0} -\alpha}$ have the same
conformal dimension and therefore can be considered as related by charge 
conjugation. Their two-point function is
\eqn\electr{\langle V_{\alpha}(\xi) V_{2\alpha_{0}-\alpha}(\xi ')
\rangle = |\xi -\xi'|^{\alpha (2\alpha _{0} -\alpha )/g}}
where $g$ is the coupling constant in \gauact . The charge neutrality 
is restored by the presence of the background charge  $-2\alpha _{0}$.
The construction of higher correlation functions involves the so called
screening operators that carry nonzero electric charge and have conformal
dimension 1. By eq. \conan\ there are two such operators which are 
charge-conjugated to each other
\eqn\scrn{\alpha_{\pm}(\alpha _{\pm}-2\alpha_{0})=4g; \ \ \ \alpha_{+}
+\alpha_{-}=2\alpha _{0}}
The allowed charges are labeled according to the number of screening charges
needed to neutralize the 4-point function \df 
\eqn\kaco{\alpha _{rs}= {1-r \over 2}\alpha _{+} + {1-s \over 2} \alpha _{-}}
The conformal dimensions of the corresponding vertex operators form the
Kac spectrum
\eqn\Kac{\Delta _{rs}= {(r\alpha _{+}/2 +s \alpha _{-}/2)^{2}-\alpha _{0}^{2}
\over 4g}}
Finally, the vortex operator with magnetic charge (discontinuity)
 $\mu$ has conformal
dimension
\eqn\dvor{ \tilde \Delta _{\mu}={g^{2} \mu^{2}/4 - \alpha _{0}^{2} \over 4g}}

The gaussian field dominance in the two-dimensional critical phenomena 
has its  microscopic equivalent. It happens that all target spaces
 with dimension not greater than one can be mapped onto the target space of 
the nonrestricted SOS model which is the discretized real line $Z$. 
The corresponding connectivity matrix
\eqn\csos{C_{xx'}=\delta _{x,x'+1} +\delta _{x,x'-1}}
has a continuous spectrum of excitations 
\eqn\ssos{\beta _{p}=2\cos (\pi p), \ \ -1 <p \le 1}
and its eigenvectors are plane waves
\eqn\esos{V_{(p)}^{x}= e^{i\pi px}}
The momentum space (the dual of the SOS target space) is therefore a 
circle with perimeter equal to two.

 The large distance behaviour of this SOS model is argued \df \gn\  to be 
described by the action \gauact\  with an appropriate choice of the
background charge $\alpha _{0}$ and the coupling constant $g$.
  The background charge in the SOS model
is introduced by taking an excited state with momentum $p_{0}$
as a vacuum state
\eqn\fvac{ S_{x}=V^{x}_{(p_{0})}=e^{i\pi p_{0}x}}
Then the curvature dependent-factor in the definition of the partition function
can be written as an exponent
\eqn\eexe{\prod_{s} \big( S_{x} \big) ^{\hat R(s)/4\pi }= \exp \Big(i{p_{0}
\over 4} \sum _{s} \hat R(s) x(s)\Big)}
which is the microscopic realization of the curvature-dependent term
in \gauact . Knowing that the global curvature does not renormalize,
we conclude that
\eqn\apap{\alpha _{0}=p_{0} + {\rm even \  integer}}
Since the charge in the SOS model is determined up to an even integer, we
can assume that $\alpha _{0} = p_{0}$.

 Now it remains to fix the coupling constant $g$ in the gaussian theory.

First we observe that
the spectrum of allowed momenta in the target space of the SOS model
is $p=mp_{0}, \ m=$integer. The operators \esos\ corresponding to momenta
outside this spectrum have vanishing correlators because the electric
neutrality is not fulfilled. On the other hand, due to 
the discreteness of the target space of the SOS model all charges of the
form $\alpha \pm 2n$ in the gaussian model are indistinguishible 
in the SOS model.
 Therefore, the spectrum of allowed electric charges in the
gaussian model is 
\eqn\alsp{\alpha = r p_{0}+2n;\ \ r,n \in Z}

The way the coupling constant $g$ depends on the background 
charge $\alpha _{0}$ is
fixed by the compatibility of \scrn   \ and \alsp 
\eqn\ccg{\alpha_{0}=g-1;\  \ \alpha _{+}=2g , \ \ \alpha _{-}=-2}
Then the spectrum of allowed charges \alsp\ coincides with
\kaco \ and the dimensions of the Kac spectrum \Kac\ read
\eqn\nkac{\Delta _{rs}={(rg-s)^{2}-(g-1)^{2} \over 4g}}
The corresponding central charge \conan 
\eqn\ccha{C=1-6(g-1)^{2}/g}
is symmetric under $g \to 1/g$.
 
By  \ccg\ the coupling constant $g$ of the gaussian field 
is related to the  ``vacuum''eigenvalue $\beta$ of the connectivity matrix
\eqn\gpg{ \beta \equiv 2\cos (\pi p_{0}) = -2 \cos (\pi g) } 
The branch of the multivalued function $g= {1\over \pi} \arccos (-\beta )$ 
 is determined by the dynamics, that is, by
the choice of the critical regime for the SOS model. The analysis of the 
$O(n)$ model \bn\ which is in a sense dual to the SOS model with $\beta = n$,
suggests that
 the interval $0<g<1$ describes
the dense phase and the interval $1<g<2$ describes the dilute phase of the 
SOS model. Sasha Zamolodchikov conjectured \ref\priz{A.B. Zamolodchikov,
unpublished} that the other branches of \gpg\  ( $2<g<\infty$)
correspond to   
 multicritical regimes   of the SOS model. This is shown to be the case 
for the SOS model on a fluctuating lattice  \ks .
 We will give a  sketch of  the analysis of ref. \ks\ in
one of the next sections.
The $m$-critical point corresponds to $g$ in the interval ($m-1,m$).

The order parameters  
 of the SOS model are represented by
 vertex operators in the gaussian theory
\eqn\orve{
 \chi _{(p)}^{x(s)}= e^{i\pi (p-p_{0})x(s)} 
\ \ \leftrightarrow \ \  V_{\alpha }(\xi )=e^{i\pi \alpha x(\xi )},
 \  \ \alpha =p-p_{0}}
The operators carrying charges $p=rp_{0}, r=0,\pm1,...$ 
have conformal dimensions in the Kac spectrum
\eqn\cdvo{\Delta^{{\rm electric}} _{(p)}= {p^{2}-p_{0}^{2} \over
 4g} ={(g-1)^{2}(r^{2}-1) 
\over 4g}=\Delta _{rr}}
The vortex operators in the SOS model can have only integer magnetic charge.
The conformal dimension of a magnetic operator with charge $\mu=m, m=
0, \pm 1, ...$ is
\eqn\dmop{ \Delta ^{{\rm magnetic}}
 _{m}= {(mg/2)^{2} - (g-1)^{2} \over 4g} =\Delta _{m0}}
The presence of negative dimensions means that the theory is not unitary.
There are two charge conjugated identity operators ($p= \pm p_{0}$).

\subsec{The loop expansion}
The partition functions of the  IRF models on a surface
 with the topology of a sphere is equal to the partition function of a
gas of nonintersecting loops on the surface \bn , \pas    .
 The loops have the meaning  of domain
walls separating domains of constant $x$. 

The microscopic construction of the loop expansion  goes as follows.
Write the partition function as a sum of monomials, choosing for each
plaquette only one of the seven terms on the r.h.s. of \wghts .
The ensemble of 
field configurations $\CS \to X$ is  divided in this way into subsets, 
each characterized by a given decoration of the plaquettes of $\CS$.
The  decorations of all plaquettes form
 a set of nonintersecting polygons on the lattice
$  \CS ^{\ast }$ dual to $\CS$ (Fig. 3).
  The polygons play the role  of boundaries of the domains 
of constant $x$.

    \epsfxsize=150pt
   \vskip 1000pt
   \hskip 80pt
   \epsfbox{ 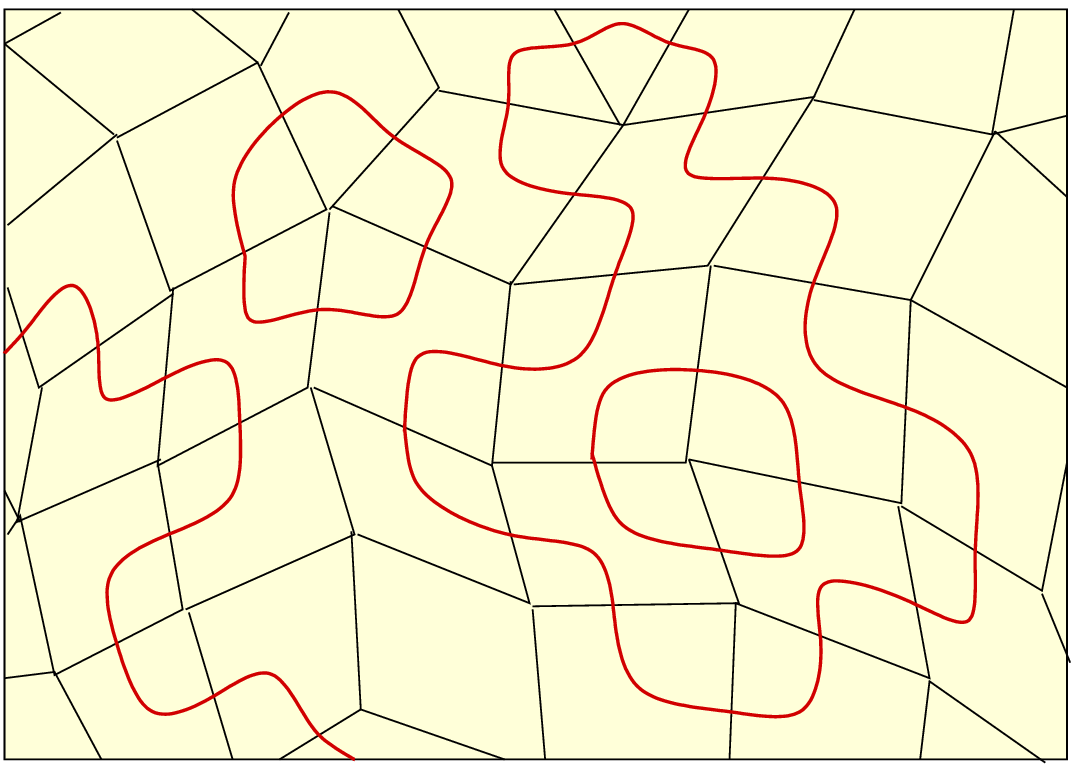   }
  \vskip 5pt
 
      \centerline{\ninepoint Fig.3:   A loop configuration
formed by decorated plaquettes.}
       
       \vskip  10pt

The partition function is equal to the sum over all configurations of polygons
on the world sheet and all allowed values of $x$ in the connected domains
bounded by polygons. The Boltzmann weight of each configuration is equal to 
$T^{-L}$ where $L$ is the total length of all polygons, times a 
product of factors associated with the domains bounded by polygons.

Let us consider for simplicity a lattice with  the topology of
a disc  with and impose a  Dirichlet boundary condition $x = constant $.
 The weight of a domain of constant height $x$
 with $n$ boundaries is $(S_{x})^{2-n}$. To prove this 
we can use the freedom to redistribute the Boltzmann weights between 
plaquettes and vertices. The new plaquette weights are given by \weightsp\
with
\eqn\nwwe{\eqalign{
 \epsfxsize=  30pt
  \epsfbox{   pqta.eps } & \qquad 
   = \delta _{x_{1}x_{3}}\delta _{x_{3}x_{4}}
\delta _{x_{4}x_{2}}\delta _{x_{2}x_{1}} 
   {1 \over S_{x_{4}}}
 \cr
\epsfxsize=  30pt
  \epsfbox{   pqtb.eps } & \qquad 
   ={1 \over T} 
  C _{x_{1}x_{3}}C _{x_{3}x_{4}}
\delta _{x_{4}x_{2}}\delta _{x_{2}x_{1}} 
  {1 \over S_{x_{4}}}  \cr
\epsfxsize=  30pt
  \epsfbox{   pqtc.eps } & \qquad =
   C _{x_{1}x_{3}}C _{x_{3}x_{4}}
C_{x_{4}x_{2}}C_{x_{2}x_{1}} 
  {1 \over S_{x_{4}}} \cr }}
and the new vertex weights are
\eqn\vrtxv{ \Omega (x) =S_{x}}
Using the fact that the number of bonds is twice the number of faces, one 
can easily show that the total power of $S_{x}$ is equal to the Euler
characteristic  $\chi =2-n$ 
of the domain. It is convenient to distribute the weight
$S_{x}^{2-n}$ as follows: a factor $S_{x}$ for the outer boundary and 
a factor $S_{x}^{-1}$ for each of the inner boundaries.

Next, for a   given configuration of
polygons we wish to sum over all possible heights of the domains bounded
by these polygons. Let us start with the domains with the topology of a 
disc, i.e., having no inner boundaries. The sum over the allowed heights
of such domain is performed using the definition \eIi . The result is
$\beta S_{x}$ where $x$ is the height of the surrounding domain.
The factor $S_{x}$ cancels  with the factor  $S_{x}^{-1}$ in the Boltzmann
weight of the surrounding domain, associated
with this boundary. Proceeding this way, from the inside out, we find that
the Boltzmann weight of a polygon configuration is obtained by assigning
to each polygon a factor $\beta = 2\cos (\pi p_{0})$. There will be also 
an overall factor $S_x$ where $x$ is the height of the most outer  domain .
 Note that the polygons have
no orientation.  One can introduce an orientation and split the weight
of a polygon into two phase factors $\exp (\pm i \pi p_{0})$ for the
two possible orientations. Note that the Boltzmann weights of the loops
do not depend on the local curvature. 

The correlation function of two order parameters, or vertex operators 
\eIv\  is equal, up to a normalization, by the partition function
of the loop gas where the  loops enclosing only one of
the two points have  different fugacity $\beta _{p}$. The correlation 
function of two desorder parameters, or vortex operators with magnetic
charge $m$ is  given by the partition function of the loop gas
in presence of $m$ nonintersecting lines having as extremities the two
points. 
We have seen that the partition function on a surface with 
the topology of a disc or sphere depends on the model only through the 
momentum $p_{0}$ of the ground
state in the target space. The mapping onto the SOS model is trivial
in this case.  For higher genus surfaces the partition
function will depend on the spectrum of allowed momenta (torus), fusion rules 
(double torus), etc. The mapping onto the SOS model then becomes more and
more involved and requires the introduction of a system of distributed
electric and magnetic charges \ref\dsz{P. di Francesco, H. Saleur, and
J.-B. Zuber, J. Stat. Phys. 49 (1987) 57}
 \fn .
  For more complicated geometries it is convenient to
formulate the loop expansion using a special diagram technique which will
be considered later in this section.
 
\subsec{Summing over the world-sheet geometries}

 Now we come to the problem of the evaluation of the string path integral.
The partition function of the string is defined formally as an integral in 
the space of all embedded surfaces .
In our case the integration with respect to the intrinsic geometry 
is equivalent to the average of the partition function in the ensemble of all
possible lattices $\CS$. The dual graphs $\CS ^{\ast }$ are generated by the 
perturbative expansion of a $\phi ^4$ matrix field theory. 
  In this approach the
length of a loop on the graph is an integer (all links of the graph are
supposed to be of unit length).  In what follows
 we prefer to consider the length $L$
as a continuous parameter, in order to simplify the notations. However,
the equivalence with the discrete formulation is complete.

For the moment both the length $L$ and the area $A$ are dimensionless
quantities. Since we are interested  in the scaling limit $A \to \infty, L \to
\infty$ , later we are going to introduce  the renormalized quantities
\eqn\cutoff{ \ell = a L,\ \ \ \CA = a^{2\nu} A}
where $a$ will be a cut-off parameter (elementary length).  It is natural
to assume that the dimension of the world sheet is 2; then $1/\nu$ gives the
fractal dimension of the loops which is determined by the dynamics.
This is also the fractal dimension $D_{B}$ of the boundary of the world
sheet with Dirichlet boundary condition on $x$.  It
can be any number between 1 and infinity. 

Let $\CS$ be any closed connected discrete manifold whose cells are squares.
Introducing the bare ``cosmological constant'' $K_{0}$ coupled to the area
 (= number of plaquettes)  $A(\CS )$ and the bare 
``string interaction constant'' $\kappa_{0}$  coupled to the
genus (= number of handles) $H(\CS )$ , we define the canonical
partition function as a sum over all such $\CS$ 
\eqn\partf{F(K_{0},\kappa _{0} )=  \sum _{\CS}  \kappa _{0} ^{2-2H(\CS )}
\ e^{-K_{0} A(\CS )} F(\CS )}
where $F(\CS )$ is the partition function for frozen geometry defined by 
\partfade .
This partition function is well defined for $K_{0}$ larger than some
critical $K_{\ast}$ where the continuum limit may exist. In what follows
the cosmological constant will be introduced implicitely through the
measure in the space of empty surfaces.

The loop expansion acquires a new significance in the case of fluctuating
geometry. It helps to reduce the problem to the problem of surfaces without 
embedding. The condition for this is the factorization of the Boltzmann
weight of a polygon configuration to a product of weights associated with 
the connected domains.  This property is very restrictif.
It excludes, for example,  terms   in the action   quadratic in the 
scalar curvature.

 The integration measure over surfaces is controlled by
two parameters: the cosmological constant $K_{0}$ coupled to the area
of the world sheet and the string interaction constant $\kappa _{0} $ 
coupled to its topology. If $F^{(H)}(A)$ is the partition function of surfaces
with fixed area $A$ and topology of a sphere with $H$ handles, then the
canonical partition function reads
\eqn\ccpf{F(K_{0},\kappa _{0}) = \sum _{H=0}^{\infty} \kappa _{0}^{H}
\int _{0}^{\infty}dA \ e^{-K_{0} A} F^{[H]}(A)}
Note that $K_{0}$ and $\kappa _{0}$ are dimensionless constants which
will be renormalized in the continuum limit $A \to \infty$. There are different
ways to achieve the continuum limit depending on the choice of the measure
in the space of surfaces.

The mapping of
 the SOS model onto a gaussian field can be constructed also in the case 
of fluctuating geometry. The sum over the geometries means functional 
integration w.r. to the metric $\hat g_{ab}$ in \gauact . In conformal 
gauge 
\eqn\cfg{\hat g ^{ab}(\xi) = \hat g^{ab}_{0} (\xi ) e^{2\pi \nu \phi (\xi  )}}
where $\hat g^{ab}_{0}(\xi ) $ is some fiducial metric,
the functional integral leads to
a theory of Liouville gravity coupled to the matter fields \ref\pols{A.M.
Polyakov, Phys.Lett. 103 B (1981) 207}.  
As has been demonstrated by David, Distler and Kawai \ddk\ , it is consistent 
to treat the Liouville field  $\phi$ as a gaussian field
 (with renormalized parameters)
and the Liouville interaction as a perturbation. The fields $x(\xi )$ and
$\phi (\xi )$ combine into a two-component gaussian field with conformal
anomaly 26 defined by the action
\eqn\actg{\eqalign{
\CA [x,\phi]& = {1 \over 4} \int d^{2} \xi \sqrt {{\rm det }\hat g _{0}}\ 
 [\pi g \ \   \hat g_{0} ^{ab} 
(\p _{a}x \p _{b}x + \p _{a}\phi \p _{b}\phi ) +  \hat  R (\xi )
 ( i p _{0}x(\xi )
-\varepsilon _{0} \phi (\xi ))
 ] \cr
& + \L \int d^{2} \xi \sqrt
 {\det \hat g_{0}} \ e^{2 \pi \nu \phi(\xi )}\cr }}  
where $\L \sim K_{0} -K_{0}^\ast $ is the renormalized cosmological
constant.

The vertex operators dressed by the fluctuations of the metric
are
\eqn\grve{V_{(p,\varepsilon )}(\xi )=e^{i\pi (p-p_{0})x(\xi)-\pi
 (\varepsilon (p) -\varepsilon _{0})\phi (\xi )}}
In particular, the puncture operator  ( =  identity operator +
 gravitational dressing) is represented by 
\eqn\idg{ \CP (\xi) = e^{-\pi (\varepsilon (p_{0})-\varepsilon _{0})\phi (\xi)}
=e^{2 \pi \nu \phi (\xi )}}
The conformal anomalies of the two components of the gaussian field are
$c_{x}=1-6 \alpha _{0}^{2}/g, \ c_{\phi} = 1+6 \varepsilon _{0}^{2}/g$
and the condition $c_{x}+c_{\phi}=26$ implies
\eqn\msco{\varepsilon _{0}^{2} - p_{0}^{2}=4g}
We choose the positive solution
\eqn\psol{ \varepsilon _{0} = g+1, \ p_{0}=g-1}
so that the two screening charges can be represented as
\eqn\scrc{  \alpha_{\pm}= p_{0} \pm
\varepsilon _{0} = g-1 \pm (g+1) \Rightarrow \alpha _{+}=2g, \alpha _{-}=-2}

The condition that the conformal dimension of the operator \grve\ is one
\eqn\didi{ \Delta _{x} +\Delta _{\phi} =
{p^{2}-p _{0}^{2} \over 4g}-{\varepsilon (p)^{2}
-\varepsilon _{0} ^{2} \over 4g} =1}
combined with \msco\ leads to the relation
\eqn\mscd{\varepsilon (p) ^{2}-p^{2}=0}
which can be interpreted as a mass-shell condition of  the same form as
 the mass-shell condition for a light particle in a two-dimensional 
space-time.  

In principle, the vertex operator \orve\ dressed by the
 fluctuations of the metric is a linear combination of two operators
\grve\ corresponding to the two solutions of \mscd 
\eqn\drv{V_{(p)}(\xi ) \to A_{+} V_{p,\varepsilon }(\xi )
+A_{-} V_{p, -\varepsilon }(\xi )}
The  two terms will have the same dimension if the coefficient $A_{-}$ contains
a positive power $a^{2|p|}$ of the cutoff $a$. Therefore the second term
is irrelevant when $p \ne 0$. Since the minimal momentum in the theories with
$C<1$ is positive $(p_{0} =1/h)$, all  physical operators correspond to
 the positive branch of \mscd 
\eqn\psel{\varepsilon (p) = |p|}
Another motivation  for the choice \psel\ is 
 based on the quasiclassical treatment
of the Liouville theory 
 \ref\seib{N. Seiberg, Rutgers preprint RU-90-29}.

In a theory with $C=1 \ (g=1, p_{0}=0)$ both terms become essential in the 
limit $p \to 0$ and the Liouville interaction in \actg\ which
 comes from the gravitational dressing of the identity operator can be not 
simply an exponential \polci  .
 
By \psol \ and \psel\ the Liouville charge of the identity operator \idg\ 
in the area term of \actg\  equals $\varepsilon _{0}-\varepsilon (p_{0})
=2\nu$ where
\eqn\nul{\nu ={1 \over 2}( g+1 - |g-1|) =\cases{
g,& if $ g<1$ ;\cr
1,&if $ g>1$ \cr } }

The gravitational dimension of the vertex operator with electric 
charge $\alpha _{rs} =p_{rs}+p _{0} =r-sg+(g-1)$ is \kpz ,\ddk 
\eqn\grdm{\delta _{rs}=1-{ \varepsilon _{0}-\varepsilon (p _{rs})
\over \varepsilon _{0}- \varepsilon (p_{0})} =
{|r-gs|-|g-1| \over |g+1|-|g-1|}}
In particular,
\eqn\mgel{\delta _{(p)}^{electric}={|p|-|p_{0}| \over 2\nu}, \ \ 
\delta _{m}^{magnetic}={mg/2 -|p_{0}| \over 2\nu }}
Finally, the so-called string susceptibility exponent $\gst $ is related to
the conformal anomaly of the matter field by 
\eqn\coco{C=1-6 {\gst ^{2} \over 1-\gst}}
It can be determined as twice the dimension $\delta _{00}$ in the Kac
spectrum of gravitational dimensions \grdm\ 
\eqn\gssg{\gst =  -2 {|g-1| \over g+1 -|g-1|}= - {\varepsilon (p_{0}) 
\over \nu } ; \ \ \ \ \ \ \nu (2-\gst )= \varepsilon _{0}}

\subsec{Diagram technique}

 We have seen that a map $\CS \to X$
 can be described by a collection of
self-avoiding nonintersecting loops dividing the world sheet into
domains of constant $x$. 
The energy of each such loop (domain wall)
 is proportional to its length $L$. In order to simplify the notations,
from now we shall consider the
length as a continuous quantity. The Boltzmann weight of a domain wall is 
\eqn\wghp{T^{-L}=\exp(-2P_{0}L)}
Further, the weight of each domain of constant height $x$ depends on the
geometry of the domain through its Euler characteristic. Indeed,
the product of the Boltzmann weights \nwwe\ associated with the
sites and squares of the domain  is
\eqn\sfcr{(S_{x})^{ {\rm sites}-{\rm squares}} =(S_{x})^{2-2H-n}}
where $H$ is the number of handles and  $n$ is the number of boundaries
of the domain.

In this way the partition function of an $ADE$ model on a surface $ \CS$ with 
arbitrary geometry has been reformulated as the partition function of a gas of 
nonintersecting loops on this surface. The Boltzmann weights of the loops 
depend on the geometry of the surface only through homotopic invariants.
In particular, they do not feel the local curvature. The loops sensible to
the topology of the surface are the noncontractible ones that wrap around
the handles of the surface $\CS$. Proceeding as in subsec. 2.3  one can check
that all contractible loops have fugacity $\beta = 2\cos (\pi p_{0})$ 
while  the fugacity of the loops wrapping around a handle is $\beta _{p}=
2\cos (\pi p)$ where $p$ is the momentum of the matter-field
 excitation propagating
along this handle. Thus the mapping of the $ADE$ models onto the loop gas 
depends, in the case of a surface with arbitrary topology, both on the 
spectrum of the order parameters and the way they interact.   
This mapping can be systematically formulated using a special diagram
technique which has been first considered in \pas\  and developed further in 
\Iade . In the case of fluctuating world-sheet geometry this diagram
technique gives rise to the Feynman rules for the corresponding string field
theory, to be discussed below.

 The sum over the embeddings of a world sheet with fixed geometry is a
formidable problem. It can be avoided by reorganizing 
the measure over embedded surfaces. First we take the sum over the world-sheet
geometries with given configuration of domain walls.  Each domain-wall
 configuration is determined by its topology and the lengths of the loops.
It can be described by a Feynman diagram with vertices corresponding
 to the domains and lines corresponding to the domain walls (Fig. 4).

    \epsfxsize=260pt
   \vskip 20pt
   \hskip 50pt
   \epsfbox{  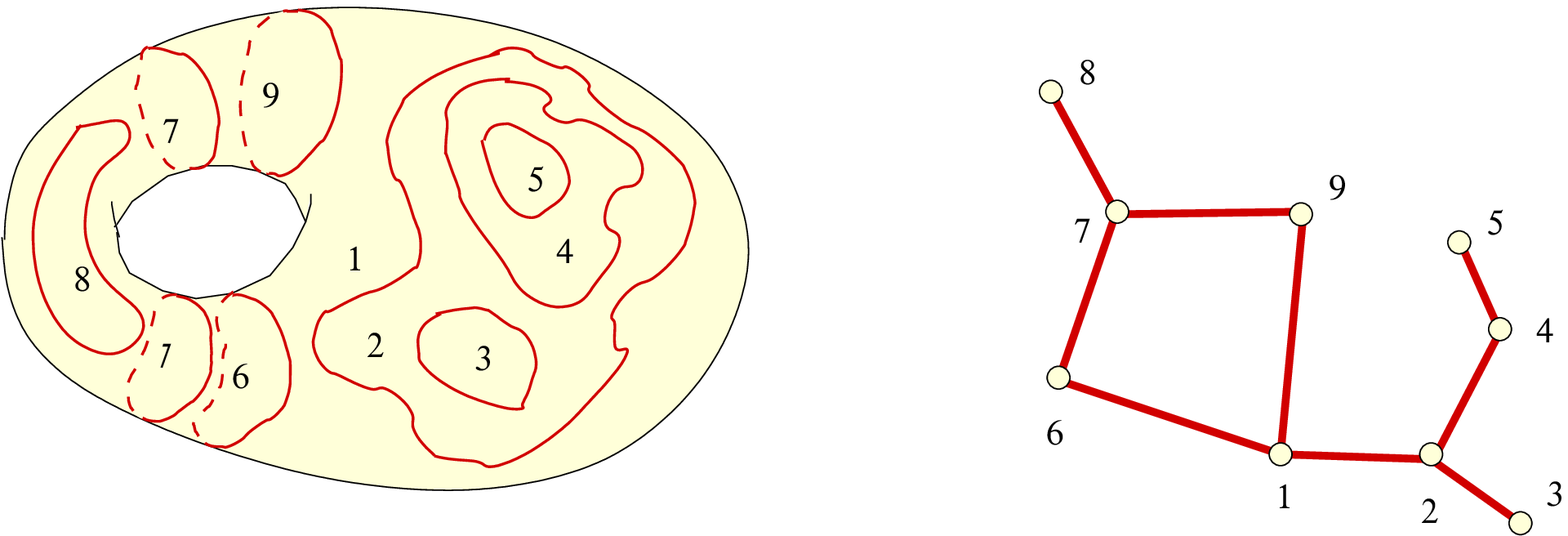  }
   \vskip 5pt
   
   \centerline{\ninepoint Fig. 4:  A loop configuration on a toroidal surface and the corresponding
Feynman diagram      }
    
    \vskip  20pt

    The heights $x$ and the lengths $L$ are associated with the vertices 
and lines, correspondingly. The partition
function \ccpf\ of the string is equal to the sum of all connected diagrams
without external lines. 

The weight of a diagram is a product of the weights
associated with its vertices and propagators. 
A vertex is defined by the coordinate $x$, the number
 of handles  $H$, and the lengths  $L_{1},...,L_{n}$ of the boundaries of
the corresponding domain. Its
 weight factorizes to a coordinate- and length-dependent components
\eqn\wever{\CV_{\c}^{(H)}(x|L_{1},...,L_{n})=( \kappa _{0}/ S_{x})^{n-2+2H}
 e^{-P_{0}(L_{1}+...L_{n})} \ 
 W_{\c}^{(H)}(L_{1},...,L_{n})}
where $ W_{\c}^{(H)}(L_{1},...,L_{n})$ is the partition function of 
nonembedded (empty
of loops) surfaces with $H$ handles and $n$ boundaries with lengths
$L_{1},...,L_{n}$.

 The ``propagator'' associated
 with a line $L$ is
\eqn\prpo{G_{0}(x,L\vert x',L')=  C_{xx'}\ \delta (L-L')} 
The integration over the lengths is performed with the measure $dL/L$.
This measure is consistent with the convention that each boundary has a
marked point on it.

It is convenient to introduce a quantum field theory in which all these 
graphs appear as  Feynman diagrams.
The latter 
 can be formulated as a functional integral over the configurations of a 
loop field $\Psi _{x}(L)$ defined in the direct product  of the target space
$X$ and the positive real axis. The interactions of
the loop field are given by the vertices \wever .

It is not difficult to see that the grand canonical partition function $Z[J]
=\exp (F[J])$ 
(i.e., the one where also disconnected surfaces are allowed) can be
written as the following functional integral
\eqn\gcpf{\eqalign{
 Z[J]=& \int D\Psi e^{S[\Psi ]}\cr
S(\Psi )=&  -{1 \over  2}\sum _{x,x'} 
\int _{0}^{\infty} {dL \over L}  \Psi _{x} (L)\big( C^{-1} \big) _{xx'}
 \Psi _{x'}(L) \cr 
&+ \sum _{x}  \sum _{n=1}^{\infty} \sum _{H=0}^{\infty} 
 {1 \over n!} \Big[ \int _{0}^{\infty} dL J_{x}(L) \Psi _{x} (L)  \cr
&+\sum _{x} \int _{0}^{\infty}{dL_{1} \over L_{1}}...{dL_{n} \over L_{n}}
\CV_{\c}^{(H)} (x|L_{1},...,L_{n}) \prod _{k=1}^{n}
\Psi _{x} (L_{k})  \Big] \cr }}
The multiloop correlation functions  are formally obtained as derivatives
with respect to the source $J(L)$
\eqn\lpsr{\langle \Psi _{x_{1}}(L_{1})...\Psi _{x_{n}}(L_{n})\rangle = 
\Bigg( \prod _{k=1}^{n} {\delta \over 
\delta J_{x_{k}}(L_{k})} F[J] \Bigg)_{J=0}}

The functional integration measure corresponds to the norm
\eqn\nrm{\|\Psi \|^{2}=\sum _{x}  \int _{0}^{\infty}{dL \over L}
\big( \Psi _{x}(L) \big)^{2}}

Formally, eq. \gcpf\ defines the string field theory corresponding to the 
target space $X$. However, in order to make this definition more explicit,
a lot of work has to be done. First, the diagram technique following from
the domain wall representation contains  tadpoles. 
This means that the string field has a nontrivial background which can
be found as the solution of the saddle-point equation
\eqn\gste{\Psi ^{cl}_{x}(L)= \Big({\delta F[J]
 \over \delta J_{x}(L)} \Big)_{\kappa _{0} =0}}
 This is
the partition function of surfaces with the topology of the
disc and bounded by a contour of length $\l$ situated at the point $x$.

It is easier to solve the classical equation of motion \gste\ after
going to  the momentum space.
For this we expand the string field as a linear combination of eigenstates \eIi
\eqn\chvr{\Psi _{x}(L)= \sum _{p} V_{(p)}^{x} \Psi_{(p)}(L)}
Let us denote by $ \CV_{\c}^{(H)}(p_{1},...,p_{n}|L_{1},...,L_{n})$ 
the vertices \wever\ in the momentum space.
They  again have  a factorized form
\eqn\pvr{ \CV_{\c}^{(H)}(p_{1},...,p_{n}|L_{1},...,L_{n})
= \kappa _{0}^{n-2+2H} N^{(H)}_{p_{1}...p_{n}}
 e^{-P_{0}(L_{1}+...+L_{n})}\  W_{\c}^{(H)}(L_{1},...,L_{n})}
where
\eqn\vrp{\eqalign{
N^{(H)}_{p_{1}...p_{n}}
&=\sum _{x} \prod _{k=1}^{n}
V _{(p_{k})}^{x} (S_{x})^{2-n-2H}\cr
&=\sum _{x} (S_{x})^{2} \chi _{(p_{1})}^{x}...\chi _{(p_{n})}^{x}
\Big( \sum _{p} \chi ^{x}_{(p)}\chi ^{x}_{(p)} \Big)^{H} \cr }}
The $p$-dependent part of the vertices 
 \vrp\ has a natural interpretation in terms of string states.
The factor $(S_{x})^{2}$ is related to the measure in the $X$ space, 
each leg is multiplied by an order parameter $\chi _{(p)}$
 and each handle  can be
considered as the result of contracting two legs
\foot{Each handle therefore contributes a factor proportional to
the volume of the target space.  In order to define the
topological expansion for the string embedded in $\Z$,
we have first to introduce a cutoff in the dual  momentum space. The simplest 
way to do this is to replace the continuum spectrum  of momenta by a discrete
one with a small spacing $\delta p$. Such a discretization appears 
automatically in the matrix-model realization of the string embedded
in $\R$ \ref\Idone{I. Kostov, Phys. Lett. 189 B (1987) 118}. The fact that
the vertices depend explicitely on the cutoff does not contradict the
general covariance. Indeed, the 
 renormalized string interaction
 constant $\kappa$  depends explicitely  on the cutoff $a$ in the space
of lenghts $\l$: it vanishes as $1/\log a$.
 If we choose $\delta p \sim 1/\log a$, then the partition
function will be cutoff-independent.}
 (in the last line of \vrp\
we have used the orthogonality relations \eIvii ).
For all target spaces 
\eqn\rva{N^{(0)}_{pp'}=\delta (p,p') ,\ \ N^{(0)}_{pp'p''} = C_{pp'p''}}
where $C_{pp'p''}$ are the fusion rules \eIvi .
 By applying several times the fusion \eIvi\  the coefficient \vrp\ can be
represented as a sum of products of fusion coefficients. For example,
\eqn\avr{ N^{(1)}_{p}=\sum _{p'}C_{pp'p'}, \ N^{(1)}_{p_{1}p_{2}}=
\sum _{p,p'}C_{p_{1}p_{2}p}C_{pp'p'},\  
 N^{(0)}_{p_{1}p_{2}p_{3}p_{4}}=\sum _{p}C_{p_{1}p_{2}p}C_{pp_{3}p_{4}}}
In general, the expression for the coefficient  $N^{(H)}_{p_{1}...p_{n}}$ 
 can be represented by a ``Feynman diagram'' with $n-2+2H$ 
vertices (to each vertex is assigned a fusion coefficient) and $H$ loops.  
Thus the p-dependent component of a multiple vertex can be decomposed into 
elementary interactions involving only three strings. Note that all ``Feynman
diagrams''representing the possible decompositions of a multiple vertex
have the same contribution (duality property).    
\foot{We believe that the $\l$-dependent component of a multiple vertex
can be decomposed as well into elementary interactions
by means of Witten's factorization arguments \ref\witt{E. Witten, Nucl. Phys.
B 340 (1990)281; R. Dijkgraaf and E. Witten, Nucl. Phys. B342 (1990) 486}}

Now we can write the functional integral \gcpf\  in  the form
\eqn\fcinp{\eqalign{
 e^{F[J]} & = \int d \Psi e^{\CA [\Psi ]} \cr
 \CA[\Psi ] & = -{1 \over 2} \sum _{p} \int _{0}^{\infty}  
\Psi _{(p) }(L)(2\cos (\pi p))^{-1}\Psi _{(p)}(L)  {dL \over L} \cr
 \ \ & + \sum _{n=1}^{\infty} \sum _{H=0}^{\infty } \sum _{p_{1},...,p_{n}} 
 {1 \over n!}\int _{0}^{\infty} 
\CV_{\c}^{(H)}(p_{1},...,p_{n}|L_{1},...,L_{n})  \prod _{k=1}^{n}
\Psi _{(p_{k})} (L_{k}) {dL_{k} \over L_{k}} \cr }}
and the classical equations of motion \gste\
 are equivalent to
\eqn\cleqm{
\Psi ^{cl} _{(p)}(L)= \delta _{p,p_{0}} W^{(0)}(L) }
\eqn\lcqm{
W(L) =\sum _{n=0}^{\infty}{\beta ^{n} \over n!}
\int _{0}^{\infty}\prod _{k=1}^{n} 
{dL_{k} \over L_{k}} e^{-P_{0}L_{k}}W(L_{k})
W_{\c}^{(H=0)}(L,L_{1},...,L_{n})}

By the loop expansion of subsec. 2.3  the quantity $W^{(H)}(L)$ is the 
partition function of the gas of nonintersecting loops on a disc with 
boundary of length $L$ and a fluctuating metric. The index $(0)$ means
planar topology of the world sheet ($H=0)$. 
Eq. \lcqm\ is graphically represented in  Fig. 5.

  \epsfxsize=80pt
 \vskip 20pt
 \hskip 100pt
 \epsfbox{   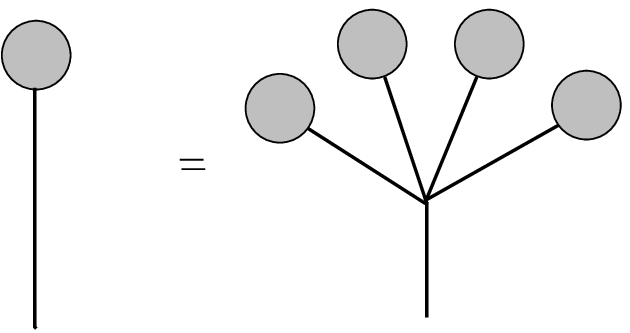}
 \vskip 5pt
 
 \centerline{\ninepoint Fig. 5:     A diagrammatic 
representation of the saddle-point equation.   }
  
  \vskip  10pt

Suppose  that we have solved the equation for the string backgroung \cleqm .
Shifting the field in the functional integral \fcinp \ 
 by its classical value
\eqn\qccq{\Psi _{(p)}(L) \to \Psi ^{cl}_{(p)}(L) +\Psi _{(p)}(L)}
we arrive at a new diagram technique without tadpoles.
The new vertices are obtained from the original ones by
 dressing with tadpoles  (Fig. 6).

   \epsfxsize=80pt
  \vskip 20pt
  \hskip 100pt
  \epsfbox{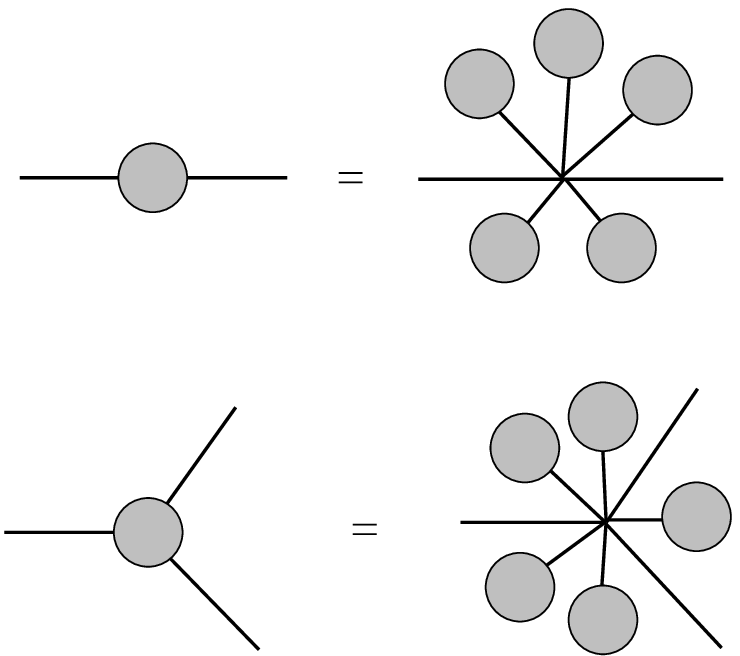     }
 \vskip 5pt
 
  \centerline{\ninepoint Fig. 6:   Graphical representation of the 
dressing of the vertices with two and three legs.     }
   
   \vskip  10pt

\eqn\vert{ \CV^{(H)}(p_{1},...,p_{n}|L_{1},...,L_{n})= 
\kappa _{0}^{n-2+2H} N^{(H)} _{p_{1}...p_{n}} W^{(H)}(L_{1},...,L_{n})}

$$ W^{(H)}(L_{1},...,L_{n})=
\sum _{k=0}^{n}{1 \over k!}\int _{0}^{\infty}
 \prod _{r=n+1}^{n+k} W^{(0)}(L_{r})
e^{-P_{0}L_{r}}{dL_{r}
 \over L_{r}}$$
\eqn\tver{e^{-P_{0}(L_{1}+...+L_{n})}
 W_{\c}^{(H)}
(L_{1},...,L_{n},L_{n+1},...,L_{n+k})}
 Note that for $n=1$ \tver\ coincides with the saddle-point equation \lcqm .
The quantity  $ W^{(H)}(L_{1},...,L_{n})$ is the partition function of the
loop gas on a fluctuating surface with $H$ handles and $n$ boundaries
with the condition that only contractible loops are allowed.

The gaussian part of the new action (the inverse propagator)  is equal to
the  sum of the inverse original
propagator  \prpo\ and the two-point
 vertex $\CV ^{(0)}(p,p'|L,L') = \delta _{pp'}W^{(0)}(L,L') $. (It is
convenient to treat the vertices $\CV^{(H)}(p_{1},p_{2}|L_{1},L_{2})$ with
 $H >0$
 as part of the interaction.) 
The propagator 
$G_{(p)}(L ,L')$ is therefore determined by the equation
\eqn\prpl{ G_{(p)}(L',L'')= 2 \cos (\pi p) \Big[ \delta (l',l'')+ 2\cos (\pi p)
 \int {dL \over L}
W^{(0)}(L',L)G_{(p)}(L,L'')\Big]}
whose solution is graphically represented in Fig. 7.

\bigskip 
   
   \epsfxsize=150pt
  \vskip 20pt
  \hskip 100pt
  \epsfbox{ 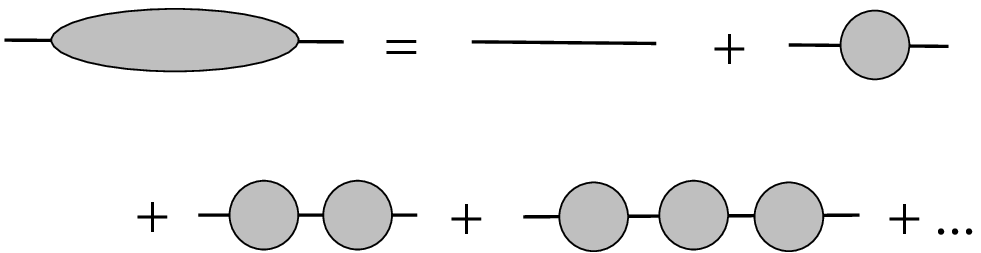   }
  \vskip 5pt
  
  \centerline{\ninepoint Fig. 7:    The full propagator
expressed in terms of two-point vertices.    }
   
   \vskip  20pt
  
    The symmetric function
 $W^{(0)}(L,L') $ 
can be considered as a kernel of a
symmetric operator in the space of the real square-integrable functions
defined on the positive real line. This operator describes the evolution of 
the string in the space of the world-sheet geometries. 
Let us introduce its eigenstates which we assume orthonormalized
\eqn\eigsts{\int _{0}^{\infty} {dL \over L} 
W^{(0)}(L,L') \langle L|E \rangle
=\Omega (E) \langle L'|E \rangle }
Then the propagator \prpl\  is diagonalized in the same basis and its
eigenstates are
\eqn\eignst{\tilde G(E,p)={ 2 \cos (\pi p) \over 1-2\cos (\pi p)\Omega (E)}}

The eigenstates $\langle L|E \rangle $ describe the excitations in the
$L$-space in the same way as the eigenstates of the connectivity matrix
describe the excitations in the $x$-space. The wave functions of the
closed string states are labeled by the ``energy-momentum'' $ (E,p)$
and have a factorized form
\eqn\factr{\langle E,p |L,x \rangle = V_{(p)}^{x} \langle E|L\rangle}

We have seen that the string background
depends on the structure of the $X$-space only through the momentum $p_{0}$
of the ground state ($p_{0}=1/h$ for the Dynkin diagrams and $p_{0}=0$
for the extended Dynkin diagrams) and therefore the same factorization 
  takes place for the vertices \vert\ describing interactions
due to joining and splitting of closed strings. This is the miracle which
makes possible 
 to solve exactly the string theory in our formalism. Note that
in our case the separation  of the two global modes occurs {\it before}
taking the continuum limit, which is not the case for the standard 
discretization of the $D=1$  string path integral
 \ref\kazm{V. Kazakov and A.A. Migdal, Nucl. Phys. B 311
 (1988) 171}. It is worth
to try to understand better the origin of  this symmetry. 

Let us mention that there exists a 
simpler model such that the only allowed momentum is $p_{0}$. This means
that the vertices have trivial coordinate part. A microscopic definition 
of this model  is given by the $O(n)$ model on a fluctuating lattice
with $n= -2\cos (\pi g)$ \ref\Ion{I. Kostov, Mod. Phys. Lett. A4 (1989)217}.

\subsec{ Loop equation}
In order to find the explicit expressions for the dressed vertices \tver\ we 
have first to solve the saddle-point equation 
 \cleqm .
 \lref\kazm {V. Kazakov, Mod. Phys. Lett. A 4 (1989) 2125}
Imagine that 
the $n$-loop amplitudes  for
 the emply random surface are constructed as the connected correlation 
functions of an $N\times N$ Hermitean random
 matrix $\Phi $ with potential $U_{\c}(\Phi )$ 
\eqn\mtam{\sum _{H=0}^{\infty } N^{2-2H} W_{\c}^{(H)}(L_{1},...,L_{n})
= \prod _{k=1}^{n} {\delta \over \delta J(L_{k})} F_{\c}[J]}
\eqn\tmam{e^{F_{\c}[J]}= \int d\Phi e^{N{\rm tr} U_{\c}(\Phi ) +
\int _{0}^{\infty } dL J(L) \tr \exp (L\Phi )}}
Then the loop amplitudes  $W^{(H)}_{\c}(L_{1},...,L_{n})$ satisfy
the following system of integral equations  \David , \Mig
\eqn\loopeq{\eqalign{
&U'_{\c}(\p /\p L)W_{\c }^{(H)}(L,L_{1},...,L_{n}) \cr
&+ \sum _{H' =0}^{H}
\sum _{I+J=\{1,...,n\}} \int _{0}^{L} dL' W_{\c} ^{(H')}
(L';L_{i}|_{i\in I})  W_{\c}^{(H-H')}(L-L';L_{j}|_{j \in J})\cr
& W^{(H-1)}(L-L',L',L_1,...,L_n)+ \sum _{k=1}^{n}W_{\c} ^{(H)}(L+L_{k};L_{s}|_{s\ne k}) =0\cr}} 
Inserting \tmam\  into \lcqm\ we arrive at the 
following closed equation for the classical string field \Icar , \Inonr
\eqn\ccg{\eqalign{
U'_{\c}(\p / \p L) W^{(0)}(L)& = \int _{0} ^{L} dL' W^{(0)}(L')
W^{(0)}(L-L') \cr
&+ \beta \int _{0}^{\infty} dL' W^{(0)}(L')W^{(0)}(L+L') e^{-2P_{0}L'}\cr}}
which should be completed with the condition that $W^{(0)}(L)$ is analytic
at the point $L=0$
\eqn\cch{W^{(0)}(L)=\sum _{k=0}^{\infty} {L^{k} \over k!} W_{k}}
In the limit $L \to \infty $ the function $W(L)$
 is expected to behave as
\eqn\asym{W^{(0)}(L) \sim L^{-b} e^{P_{R}L}}
where the entropy per unit length $P_{R}$ depends on the parameters of
the theory.

Eq. \ccg\ has a transparent geometrical meaning \kazm . A small deformation of
the world sheet at the marked point is equivalent to a local deformation of
the boundary. Its exact form depends on the measure in the space of the
world-sheet geometries.In general, a local deformation of the 
boundary is equivalent to a differential operator
\eqn\potential{ U_{\c}'(\p /\p L) =
 g_{2} \p /\p L + g_{3} \p ^{2} /\p L^{2} +...}
On the other hand, the deformation of the world sheet cancels part of the
integration measure in the space of surfaces and produces boundary (or
contact) terms. The standard  contact term, which appears also in pure
gravity, is due to the surfaces with degenerated world sheet
such that another point of the boundary approaches the marked point. Here
we encounter a new contact term due to one of the loops approaching the
marked point at the boundary.

The loop equation \ccg\ has been derived originally \Iade
in terms of the coefficients $W_{k}$ 
\eqn\eem{\eqalign{
&\sum _{k\ge 2} 
 g_{k} W_{n+k-2} =\cr
&\sum _{k=0}^{n-2}W_{k}W_{n-2-k} + \beta \sum _{p,q=0}^{\infty}
 {(p+q)! \over p!q!}   e^{-2P _{o}(p+q+1)} W_{p}W_{n+q-1}\cr }}
We arrive at this form of the loop equation if the gas of loops is considered
on a random graph, the length of the loops being the number of bonds 
they occupy. The formulation in terms of continuous lengths has been proposed
by Kazakov \kazm .

%

It is convenient to introduce the Laplace image
\eqn\ccj{ \hat W(P)= \int _{0}^{\infty} dL \  W^{(0)}(L)e^{-PL}}
which can be interpreted as the partition function of 
surfaces with the topology of a disc and ``boundary cosmological constant''
$P$. In order to simplify the notations we do not assign an index $(0)$
to $\hat W(P)$. The condition of analyticity  \cch \ implies
\eqn\cck{\hat W(P)={W_{0} \over P} +{W_{1} \over P^{2}}+..., \ \ \ 
P \to \infty}
(In what follows we assume the standard normalization $W_{0}=1$.)
The asymptotics \asym\ means that the series \cck\ 
converges for $|P|>P_{R}$; in the vicinity of the point
$P_{R}$ it behaves as $(P-P_{R})^{b-1}$.

We are interested in the solutions of the loop equation having a single
cut $[P_{L},P_{R}]$ along the real $P$-axis. The 
loop equation \ccg\ in terms of $P$-variables reads
\eqn\ccl{\hat W^{2}(P)=\oint _{\CC} {dP' \hat W(P') \over 2\pi i (P-P')}
[- U_{\c}'(P')-\beta \hat W(2P_{0}-P')]}
where the contour $\CC$ encloses the cut $[P_{L},P_{R}]$ of the Riemann 
surface of the function $\hat W(P)$. The contour integral makes sense 
only if the two cuts $[P_{L},P_{R}]$ and $[2P_{0}-P_{R},2P_{0}-P_{L}]$ of the
Riemann surface of the integrand function do not overlap, that is, for 
$P_{0}>P_{R}$.The positions 
$P_{L}$ and $P_{R}$ of the branchpoints are functions
of the potential $U_{\c}(P)$ and the temperature $P_{0}$ of the loop gas.
Eqn.  \ccl\  implies the following conditions on the 
real and imaginary part of $\hat W(P)$ \Iade  
\eqn\ffr{\eqalign{
2 {\rm Re} \hat W(P) + \beta \hat W(2P_{0}-P) +
U_{\c}'(P)=0, &P \in [P_{L},P_{R}] \cr
{\rm Im} \hat W(P)=0 , & P \not\in [P_{L},P_{R}] \cr }}

\newsec{ Critical behaviour }

\subsec{Critical surfaces, noncritical loops}
The partition function \partf\  is defined in certain domain $\CD$ in the
space of the parameters $P_{0},K_{0}$,... . The critical behaviour is
achieved at the boundary $\p \CD$ of this physical domain, where the partition
function and the observables develop singularities.
There are  two types of singularities in the space of parameters.
The first is related to the diverging area of the 
fluctuating surface. It appears when the coupling constants $g_{1}, g_{2},...$
 in the potential $U_{\c}(P)$ which defines the integration measure in
 the space
of nonembedded surfaces, are tuned in a special way. 
The loops on the world sheet
 are small and do not affect the scaling properties.
The critical behaviour of the model is the (multicritical)
behaviour of pure gravity. The $m$-critical points of the nonembedded random 
surface \kazm\ have been interpreted \stau ,\bdks , \gm , \ref\cgm{ C.   
Crnkovic, P. Ginsparg, and G. Moore, Phys. Lett. 237 B (1990) 196}
as pure gravity coupled to a conformal field theory $(2,2m-1)$. It 
can be mapped onto the Coulomb gas \actg\ with $g=m-1/2$.
The $m$-critical potential has been chosen in \kazm\ as an even polynomial
of degree $2m$
\eqn\mcpt{U_{\c}(P)=\sum (-)^{k-1} {(k-1)!m! \over (2k)! (m-k)!}P^{2k}}
However, if the symmetry $P \leftrightarrow -P$ is abandoned, it suffices to
take a polynomial of degree $m+1$. The scaling limit
 is achieved by replacing $L,P$ by the renormalized quantities $\l, z$
\eqn\renm{L=\l /a, \ \ P=P_{\ast} + az}
 At the $m$-critical point the scaling part $\hat w(z)$ of the planar 
loop amplitude is defined 
by 
\eqn\lili{\hat W(P)=\hat W(P_{\ast}) + a^{m-1/2} \hat w(z)}
In the matrix-model realization of the multicritical points the renormalised 
cosmological constant $\lambda $ scales as $a^{-m}$ which could mean that the
fractal dimension of the boundary of the world sheet is $2/m$. For $m>2$
this is less than the classical dimension 1. The resolution
of this paradox comes from the observation \stau \bdks that for
$m>2$   $\lambda $
is coupled to the operator with minimal (negative)
 gravitational dimension $\delta _{m1}=1-m/2$
and therefore does not measure the area of the world sheet.
The true cosmological constant $\L$ has dimension $a^{-2}$ for $m\ge 2$
and $a^{-1}$ for $m=1$. Finally, the
string interaction constant scales as
\eqn\srta{\kappa _{0}=\kappa \ a^{1-\gst /2}; \ \ \gst = 3/2 \ -m}

\subsec{Dense phase}

When $P_{0}$ approaches its critical value $P_\ast$, long-range effects
due to the diverging size of the loops change the critical behaviour of
the model. For generic potential $U_{\c}$ the random surface
grows only due to the critical  loops which are densely packed
on the world sheet. Its area is essentially equal to the total
length of the loops. This is the {\it dense} phase of the loop gas. If both
the potential $U_{\c}$
 and the temperature $P_{0}$ are tuned, a more complicated picture arises.  

The regime of critical loops coupled to noncritical surfaces describes
the {\it dense phase} of the loop gas model. In this regime the loops are 
densely packed and fill almost all surface of the world sheet. The 
nonrenormalized area
of the world sheet is therefore equal to the total (nonrenormalized) length
of the loops: $A=L_{{\rm tot}}$
and $P_{\c}$ plays the role of a cosmological constant.

It is easy to see that the condition for complete compensation of the entropy
and the energy of the loops is  $P_{R}=P_{0}$ . First let us remind that
$W(L)$ behaves when $L$ is large according to \asym .
Let us choose one of the random loops on a surface with the
topology of a sphere. It splits the sphere into two
discs, each contributing a factor $e^{P_{R}L}$ when $L$ goes to infinity. Thus
the entropy $2P_{R}L$ of the loop totally compensate its energy $2P_{0}L$
when  $ P_{R}=P_{0}$.

The vicinity of the critical point is defined by introducing a cutoff
parameter $a$ with dimension of length. The variable  $z$ dual to the
renormalized length $\l$  (the renormalized boundary cosmological constant) 
is introduced by
\eqn\crtp{P=P_{\ast}+az, \ \ P_{R}=P_{\ast}-aM}
The parameter $M$ is the contribution of the fluctuations
 of the surface to the renormalized boundary cosmological constant.
It defines the position of the cut in the $z$-plane.

The singular part of the loop amplitude $W(P)$ behaves at the critical point
$P_{R}=P_{0} \ \ (M=0)$ as $z^{g}$ where $g$ is the solution of $2\cos 
(\pi g) +\beta =0$ , $0<g<1$ \ref\gk{M.
Gaudin and I. Kostov, Phys. Lett. B220(1989)200},\Iade.
 Therefore we define the scaling part
$\hat w(z)$ of the loop amplitude as
\eqn\sccc{\hat W(P)=\hat W_{\ast} + ({\rm constant}) a^{g}\hat w(z)}
where $\hat W_{\ast}$ is the value of the bare loop amplitude for 
$P=P_{\ast}, P_{0}=P_{\ast}$.

Now we are going to find 
 the scaling law for the cosmological constant $\Lambda$ which is defined as
\eqn\coct{P_{0}=P_{\ast}- ({\rm constant })a^{2\nu }\L}
 If we assume that the dimension of the world sheet is 2, then
the exponent $\nu$
has the meaning  of inverse
 fractal dimension of the boundary of the world sheet.
The induced boundary cosmological constant $M$ is a function of $P_{0}$ 
which vanishes at $P_{0}=P_{\ast}$.
 Therefore the renormalized bulk and boundary
 cosmological constants are related by
\eqn\hdimb{\Lambda = ({\rm constant }) M^{2\nu}}

The scaling of $\L$ can be extracted from a nonlinear algebraic equation
which follows from the loop equation \ccl .  In order to get rid of 
the contour integration we use the following trick.  First write
again eq. \ccl   \  with the variables
 $P,P'$ replaced by their images w.r.t. the
reflection $P \to 2P_{0}-P$.  The integrand remains the same but the
contour $\CC$ is replaced by a contour $\bar \CC$ enclosing the cut
$[2P_{0}-P_{R},2P_{0}-P_{L}]$ of $\hat W(2P_{0}-P)$. Then add the two 
equations and apply the Cauchy theorem to the integral along the contour
$\CC +\bar \CC$. The result is the following functional equation for
$\hat W(P)$
\eqn\frty{\eqalign{
& [-U'_{\c}(P) \hat W (P) - \sum _{k \ge 0} P^{k}(g_{k+2} W_{0}
+g_{k+3}W_{1}+...)  -\hat W^{2}(P)] \cr
& +[P \to 2P_{0}-P] = \beta \hat W(P) \hat W(2P_{0}-P)\cr}}
At the point $P=P_{0}$ this functional equation becomes algebraic
\eqn\algb{(2+\beta)\hat W ^{2}(P_{0}) +2 U'_{\c}(P_{0})\hat W(P_{0})
-2 \sum _{k \ge 0} P_{0}^{k} (g_{k+2} +g_{k+3} W_{1}+...)=0}

Eq. \algb \ becomes very simple in the limit $P_{0} \to  -\infty$ where
the only allowed loop configurations are those with no space between loops.
This extremely dense limit can be achieved by choosing 
 gaussian potential $(g_{2}=-1, g_{3}=...=0)$. Then eq. \algb\  involves only
the coefficient $W_{0}=1$ of the expansion  \cck\  and can be solved instantly
\eqn\gsss{\hat W(P_{0}) = 2/(P_{0}+\sqrt{P_{0}^{2}-2(2+\beta )})}
This quantity has a singularity at  $ P_{0}= \sqrt { 2(2+\beta)}
 =P_{\ast}$ in accord with the exact  result obtained with  more
elaborated technique \gk .

By the definitions  \sccc\  and \coct\ we find 
\eqn\ssc{W(P_{0})=W_{\ast} + ({\rm const.})a^{g}\hat w(0)
 =W_{\ast}-({\rm const.})\sqrt {\L}}
By its dimension $\hat w(0) \sim M^{g}$ and eq. \ssc\ implies $\nu =g$.
>From now we will fix the normalizations of $\L$ and $M$ so that
\eqn\nrmr{\hat w(0) =- \cos (\pi g /2) M^{g}, \ \ \L=M^{2g}}
(Note that the sign of $ \hat w(M)$ should be negative.)
 This scaling law persists for  generic potential $U_{\c}(P)$,
i.e., in a domain $\CD _{1} $ of codimension zero in the space of coupling
constants $g_{2},g_{3},...$ . Indeed, let us repeat the above argument 
for a generic potential $U_{\c}(P)=g_{2}P^{2}+g_{3}P^{3}+...$.
Expanding eq. \algb\
 around the point $P_{0}=P_{\ast}$ we find a relation
between $\hat w(0) =- \cos (\pi g /2) M^{g}$ and $\Lambda$. 
In the limit $a \to 0$
\ the most singular part of the coefficients $W_{k}$ is the v.e.v. of the 
puncture operator $\CP = -\p /\p \Lambda $
\eqn\punc{ W_{k} \sim a ^{2\nu (1- \gamma _{\rm str})} \langle \CP \rangle 
\sim (a^{2\nu}\Lambda )^{1-\gamma _{\rm str }} }
Inserting this in \algb\ we find a relation of the type
\eqn\stiga{(aM)^{2g} = \sum _{k \ge 1} A_{k}(a^{2\nu} \L)^{k} +
B(a^{2\nu}\Lambda )^{1-\gamma _{\rm str}}}
where $A_{1},A_{2},... $ and $B$ are numerical coefficients. 

In the limit $a \to 0 $ only the smallest power in $a$ will survive.
For a generic potential all the coefficients $A_{k}$ are nonzero.
The strongest singularity is that of the $A_{1}$-term and the matching of 
powers gives
\eqn\powers{2g = {\rm min}[2\nu, 2\nu (1- \gamma _{\rm str})]}
Sinse the string susceptibility exponent is allways non-positive, this
implies $ \nu = g$.
In order to fix the string susceptibility exponent, we use the fact that 
it is related to the dimension of the loop amplitude
\eqn\sledpolunost{\nu (2-\gamma _{\rm str})-1=g}
Thus in the dense phase 
\eqn\crex{\nu =g,\ \ \gst  =1-1/g; \ \ \ 0<g<1\ \ \ \ \ \ ({\rm dense\ 
 phase})}
In the interval $1/2 <g<1$ the difference $P_{\ast} -P_{0}$ vanishes
faster than $a$ and the point $P_{0}$ in the definition of the scaling
variables can be replaced with $P_{\ast}$. This is not the case, however, 
if $g<1/2$. In this case the cut of the loop amplitude appears at
$z=-M-a^{2g-1}\Lambda $ and in the
limit $a \to 0$ the singularity of the loop amplitude is gone to 
$-\infty$.  We have a very poor understanding of this case at the moment. 
Let us only mention  that the fugacity of the loops is negative and the 
fractal dimension of the boundary is larger than two. The points $g=1/m, \
 m=1,2,3,...$ have integer $\gst $ and should describe topological theories.
A very interesting discussion on the physics at these points has been
given by H. Saleur \ref\sal{H. Saleur, Chicago preprint EFI
90-92, December 1990;  Nucl. Phys. B 360 (1991) 219 }
 (see also \ks ).

The limiting  case  $g=1/2$ corresponds to the gaussian model. In this case
there is no singularity at all and the choice of the distance $M$ is 
arbitrary.
 
\subsec{Dilute phase}

Now let us consider the case when both kinds of singularities are present.
If the potential
$U_{\c}(P) $ is critical $ (m=2)$ , then at the point $M = M_{\ast }$ both the
length of the loops and the area of the space between them become infinite.
This phase is known as the {\it dilute phase } of the loop gas.
The multicritical potentials will lead,further, to new
types of critical behaviour of the gas of loops \ks \ which we are going to
discuss briefly.

As we have seen above, the dense phase exists in a domain 
$\CD _{1}$ of codimension zero in the space of coupling constants $g_{1},
g_{2},...$. The generic point on the boundary $\CD_{2}=\p\CD_{1}$ of this
domain corresponds to the next critical regime. The simplest way to
achieve it is to adjust the constant $g_{3}$ in a cubic potential
so that the coefficient $A_{1}$ in \stiga\ vanishes. Then the condition of
matching of powers reads
\eqn\critgr{2g ={\rm min}[4\nu,\ 2\nu (1- \gamma _{\rm str})]}
If we are in the interval $1<1-\gst <2$, this condition 
means $2g=2\nu (1- \gst )$ . Then from \sledpolunost\ we find
\eqn\utre{ \nu = 1, \ \ \gst = 1-g ;\ \ \ 1<g<2  \ \  ({\rm dilute\  phase})}
Proceeding in the same way , we can achieve the $m$-critical point by 
adjusting the coupling constants $g_{3},...,g_{m+1}$ so that the 
coefficients $A_{1}, ..., A_{m-1}$ vanish. The condition of matching of 
powers then gives \ks 
\eqn\mcritgr{2g={\rm min}[2m\nu , 2\nu (1- \gst )]}
Combining \mcritgr\ with \sledpolunost\ we find
\eqn\crexm{\nu =1,\ \ \ \gst = 1-g ;\ \ \ m-1 
<g<m \ \ \  (m { \rm  -critical \  dilute \ phase})}
Let us assume that the loop gas with given $g$ describes at the critical 
point a conformal field theory coupled to gravity. Then, by the
KPZ-DDK argument  \kpz , \ddk \       its conformal anomaly is given by 

\eqn\centrch{C=1- 6 {\gst ^{2} \over 1-\gst } = 1- 6{(g-1)^{2}\over g}}
Thus the interval $0<g<\infty$ covers twice the spectrum $-\infty <C<1$
of the central charge. Note that even if the central charge is symmetric
w.r.t. $g \to 1/g$, the fractal dimension of the boundary is not.
In the dilute phase it takes the ``classical'' value 1 and in the 
dense phase it is greater than one. 

 The $m$-critical quantum gravity coupled to a gas of critical
loops exhibits a specific critical behaviour characterized by the fugacity
 $\beta$ of the loops. One can say that the $(2m+1,2)$-conformal theory of 
matter coupled to the gas of loops with fugacity $\beta$  produce an
effective matter field with central charge $C=1-6(g-1)^{2}/g$ where $g$ is
the branch of the function $g=-(1/\pi) \arccos ( \beta /2)$ 
determined by $m-1 <g<m$.

The change of the critical regime occurs at the integer points $g=m, \ 
m=1,2,3,...$. Let us consider the vicinity of the point $g=m$
dividing the $m$-critival and $m+1$-critical dilute phases of the loop gas.
In this case eq.\sledpolunost\ reads
\eqn\beraha{ M^{2}= \L [A_{m} (a^{2}\L )^{m-g} +B]^{1/g}}
The first term on the r.h.s. vanishes when $g<m$ because the power of $a$
 is positive as well as when $g<m$ because then $A_{m}=0$. However, when $g=m$
both terms survive and   our previous arguments may not be applicable. The
limits $g \to m $ and $a \to 0$ do not commute and one has to solve the
loop equation for $g=m$ and {\it then} take the continuum limit $a\to 0$.
Exact solutions are known for the cases $g=1$ 
\ref\godu{M. Gaudin, unpublished} and $g=2$ \ref\km{I. Kostov and M. Mehta,
Phys. Lett. B189 (1987)118}, to be discussed below. At $g=2$ the the 
scaling is the same as in the whole dilute phase, $\L =M^{2}$. At the
point $g=1$ which is the endpoint of the dilute phase, the scaling law 
receives  logarithmic corrections: $\L = (M \log M )^{2}$.
 
%

\subsec{The loop equation and the string background in the continuum limit}

We are going to solve  the loop equation \ccl\ in the continuum limit.
 Let us define the vicinity of the critical point as follows
\eqn\critp{\eqalign{
P&=P_{0}+az \cr
P_{R}&=P_{0}-aM \cr
\hat W(P)&=(2U'_{\c}
(P_{0}+az)- \beta U'(P_{0}-az))/(4-\beta ^{2})+\ ({\rm const.})
 a^{g}\hat w(z)}}
Here we used as a reference point $P_{0}$ instead of $P_{\ast}$; we shall
see that if $1/2 <g<1$ this makes no difference in the limit $a \to 0$.

After  the vicinity of the critical point is blown up by the change of
variables \critp\ , the cut $[P_{L},P_{R}]$ of the Riemann surface of $W(P)$
is replaced by the cut $[-\infty,-M]$ along the negative $z$-axis.
Eq. \ffr \ implies the following conditions on the real and imaginary
 parts of $\hat w(z)$
\eqn\scalir{\eqalign{
2{\rm Re} \hat w(z) + \beta \hat w(-z) =0, \ \ \  &  z\le -M \cr
{\rm Im}\hat w(z)=0     , \ \ \  & z \ge -M \cr }}
Eq \scalir\ becomes more transparent when written in terms of the variable 
$\tau$ which we have introduced in \Icar ,\Inonr
\eqn\parr{z=M\cosh (\tau )}
Then the $z$-plane cut along the interval $-\infty <z<-M$ is mapped 
into the half-strip 
\eqn\ffz{ {\rm Re }\tau \ge 0, -\pi \le {\rm Im }\tau \le \pi}
The two sides of the cut are mapped to
the horizontal 
boundaries [$\tau = t \pm i \pi , t>0$] of the $\tau$-strip. Therefore ,
if $z$ is real and positive, then $-z \pm i0 = M\cosh (\tau \pm i\pi )$.
Since $\hat w(z)$ is a real function ( $(\hat w(z))^{\ast} = \hat w(z^{\ast})$), the first of the eqs. \scalir\ can be written as
\eqn\ham{[\cos (\pi \p /\p \tau )- \cos (\pi g)]\hat w(z(\tau ))=0}
Eq. \ham\ has a single solution (up to a constant factor) 
 which is real along the vertical 
boundary of the $\tau$-strip and behaves  at infinity as  $z^{g}$
\eqn\solu{\hat w(z)=-M^{g} \cosh (g\tau )}

If we return to the original variable $z$, the solution reads
\eqn\ffx{\hat w(z)= - {1 \over 2}
[(z+\sqrt{z^{2}-M^{2}})^{g} +(z- \sqrt {z^{2}-M^{2}})^{g}]}
It can be expanded
 as an infinite series in fractional (in general)  negative  powers
of $z$ with radius of convergence $1/z =1/M$
\eqn\ffv{\hat w(z)=\sum _{n \ge 0} w_{n}^{\pm } M ^{g} (M /z)^{2n \mp g}}
and dimensionless coefficients
\eqn\ffw{\eqalign{
w_{n}^{\pm}& = \pm {\Gamma (2n \pm g) \over \Gamma (n+1\pm g)n!}\cr 
&= \pm {(2n \pm g)(2n \pm g-1) ... (n+2 \pm g) \over n!}\cr }}
Even if each of the terms in \ffv\ has a cut $-\infty <z<0$, the whole
series  defines a function which is analytic  for $|z|<M$.

The v.e.v. of the loop operator is determined by the same equation \scalir  \  
both in the dense and dilute phases. It is given by the same analytic
function \ffx  \  with $g$ ranging from 0 to  $\infty$. The $m$-critical 
behaviour of the loop amplitude is described by the branch 
$m-1 <g <m$  of the parametrization  \gpg .  
The function $\hat w(z)$ has a square-root singularity at $z=-M$ . It is
meromorphic in the $z$-plane with a cut from $z=-M$ to $z=-\infty$ where
it behaves as $z^{g}$. For nonrational values of $g$ the Riemann surface of
$\hat w(z)$ is infinitely foliated and has two cuts $-\infty <p<-M$ and $M<p
<\infty$. All sheets except the first one have two cuts. We will
 always consider the function $\hat w(z)$ on the first sheet.   For
 $g=p/q$  the Riemann surface of $w(z)$
has a branch point of order $q$ at infinity. Then the first
and the last sheet have only one cut.

If we introduce a renormalized length $\ell=aL$, 
then the inverse Laplace image of $\hat w(z)$ 
\eqn\fffc{w(\ell )=\int _{-i\infty}^{i\infty} dz e^{\ell z}\hat w(z);
\ \ \ \hat w(z) = \int _{0}^{\infty} d\ell \ e^{-z \ell }w(\ell)}
is the Bessel function \mss
\eqn\fffd{ w(\ell) =  {M^{g} \over\ell } K_{g}(M \ell).
}
It  satisfies the following loop equation
\eqn\lqaa{\int _{0}^{\infty} d\l ' w(\l ')w(\l -\l ')+
\beta \int _{0}^{\infty} d\l ' w(\l ')w(\l + \l' ) \ =0}
which is the renormalized version of \ccg .
The renormalized  string background \fffc\ 
behaves differently at large and small lengths
\eqn\sssspace{
w(\l)= \l ^{-1} M^{g}K_{g}(M\l ) \sim 
\cases{ 
M^{g-1/2} \l ^{-3/2}e^{-M\l }
,  & if $ \l \gg M^{-1} ;$\cr
 \l ^{-g-1} ,& if $\l  \ll M^{-1}$\cr     }}
The small-$\l$ asymptotics \sssspace\ can be taken as a subsidiary condition
to eq. \lqaa\ in order to have a unique solution.
The large-$\l$ asymptotics describes a loop of length 
much larger than the characteristic
length $\bar \l =1/M$ . The area of the world sheet is zero in this limit 
and the power of $\l$ is the same as in the topological gravity $(g=1/2)$.
The small$-\l$ asymptotics describes the critical point where the area of the
world sheet and the characteristic length of the loops are infinite.
We see that the square-root singularity changes generically
 to a nonrational one in 
the limit $M \to 0$.

In order to  to reproduce the regime of critical surfaces, noncritical 
loops we have to take the limit $\L \to 0$ ,  keeping $M$ finite,
 i.e., to leave the trajectory $\L = M^{2\nu}$.
 However, we can achieve the critical points of nonembedded surfaces 
by tending the fugacity $\beta $ of the loops to zero. This limit is
achieved at the  half-integer values of $g$. 
By the scaling of the loop amplitude $\hat w(z)$ these points can be identified with the multicritical points of pure gravity \kazm .
The explicit expression for the loop average  $\hat w(z)$  
in the $m$-critical point is
\eqn\ffk{
\hat w(z)= -\sum_{k=0}^{m} (1+(-)^{k})2^{m-1/2} {(2m-1)! 
 \over k! (2m-1-k)!}(z+M)^{m-1/2-k/2} (z-M)^{k/2}}
Eq. \ffk\ reproduces the known results (see, for example, \David)
 for $g=3/2 $ (pure gravity) and $g=1/2$ (gaussian model) 
\eqn\odve{\hat w(z)=
\cases{
-2(2z-M)\sqrt{(z+M)/2}, & \ $ g=3/2$ \cr
-2\sqrt{(z+M)/2} , & $ \ g=1/2$ \cr }}

At the integer points $g=m, \ m=1,2,3,4,...$ 
the planar loop amplitude \ffx\  is a polynomial
and has  no cut at all. However,this is not always
 the physical solution because
the limit $a \to 0$ is taken prior to the limit $g \to m$. The exact solution
for the case  $g=2$ \km\  lead to the following expression
in the continuum limit \foot{ The equivalence between the gas of dilute
loops with fugacity $\beta =-2$ ($g=2$) and the bosonic string embedded in
-2 dimensions can be established as follows. If the loops are considered as 
oriented, then their fugacity  is -1. Further, it is easy to see that the
condition of nonintersection can be abandoned, since the total contribution 
of the configurations containing intersecting loops vanishes because of
cancellations. Therefore the partition function can be written as the
exponential of minus the entropy of a single oriented loop, with no
restriction to its configurations. This is exactly the determinant of the 
Laplace operator in the lattice $\CS$, which can be written as a  Gaussian
integral with respect to a couple of grassmanian fields $x,\bar x$ 
defines on $\CS$.
Taking the derivative w.r. to $g$  we subtract the zero mode
of the Laplacian.
By the Kirchoff theorem, the finite part of the determinant   is equal to the
sum of connected spanning trees on the world sheet. Since each spanning
tree defines a dense loop on the world sheet \ref\dud{B. Duplantier and
F. David, J. Stat. Phys. 51 (1988) 327}, the derivative of
the partition function can 
be also expressed in terms of the gas of dense loops with fugacity $\beta 
\to 0$.
We see that the {\it derivatives} of the models $g=1/2$ and $g=2$  are
identical and describe the string in -2 dimensions.  }
\eqn\gone{\eqalign{
\hat w(z) _{g=2} & = -\sum _{+,-} \Big(z \pm \sqrt {z^{2}-M^{2}}\Big) ^{2}
\log  \Big(z \pm \sqrt {z^{2}-M^{2}}\Big)\cr
&= \Big[{d \over dg}\hat w(z) \Big]_{g=2},
\ \ \ \ \ \ \L=M^{2} \ \  \cr}}

The case $g=1$ has been solved by M. Gaudin  \godu \ for the gaussian 
potential  $U_{0} = -1/2 \Phi ^{2}$. The imaginary part of $\hat w(z)$
along the cut $- \infty <z<-M$  has been  found as the solution of an integral 
equation  with Cauchy kernel.  From this it is not difficult to reconstruct 
the meromorphic function $\hat w(z)$
\eqn\novo{\eqalign{
\hat w(z)_{g=1} &= -M[\tau ^{2} \cosh \tau + 2 \tau  \log (aM)
\sinh \tau]  \cr
&=-2 \sqrt{z^{2}-M^{2}}
\log (aM) \log \Big({ z+\sqrt{z^{2}-M^{2}} \over M} \Big) \cr
 & -z \Big[\log \Big({ z+\sqrt{z^{2}-M^{2}} \over M} \Big)\Big]^{2} \cr
&= \Big[{d^{2} \over dg^{2}} \hat w(z) \Big]_{g=1} + (const) \ z, \ \ \ \ \ \ 
\L  = [M \log (aM)]^{2} \cr }} 
The origin of the logarithmic corrections  in these two cases is quite
different. In the case $g=1$ this is the appearance of a massless
excitation (the ``tachyon''). The divergences due to this zero mode
produce the logarithmic factor in the relation between $\L$ and $M$.
The logarithmic scaling violation  is a well known phenomenon in the 
string theory with $X=\R$  \kazm , \ref\bkz{E. Br\'ezin, V. Kazakov and
Al. Zamolodchikov, Nucl. Phys. B 338 (1990) 673;
 D. Gross and N. Miljkovic, Phys. Lett. 238 B (1990) 217;
 P. Ginsparg and J. Zinn-Justin, Phys. Lett. 240 B (1990) 333; 
G. Parisi, Phys. Lett. 238 B (1990) 209}.
In the case $g=-2$ there are no divergences which can produce logarithmic
corrections in the scaling of $M$.  The logarithmic factors in the loop 
amplitude are related to the fact that the partition function vanishes
and therefore the loop amplitude is actually a {\it derivative} \sal .
Within the matrix-model representation, the vanishing if the partition
 function can be explained with the occurence of a 
supersymmetry at the point $g=-2$ \ref\dvid{F. David, Phys. Lett. B 159 (1985)
303},\km . This has nothing to do with the sum over geometries.
 The partition function for fixed world-sheet geometry
 is equal to the determinant of the Laplace operator on the corresponding
lattice and vanishes due to its zero mode.

As for  the higher integer points, $g=3,4,...$, it is possible that there are
no logarithmic singularities \foot{Saleur \sal\ considered the topological
points $g=1/2, 1/3,...$ which are in a dual to the points $g=2,3,...$.
He argued that a self-consistent definition of the theory as a derivative
in $g$ exists only for $g=1/2$.}
These points will be discussed in \ks .

We would like to understand why the loop amplitude of the $g=1$ string
is equal to the second derivative in $g$ of the generic solution. For the
moment this is just an experimental fact.

 \newsec{The multi-loop amplitudes in the scaling limit}

\subsec{An effective matrix model for the dressed vertices}
In order to be able to exploit  the functional integral formulation \gcpf   \ 
of the $ADE$ and SOS strings  we need explicit expressions for the
 vertices \vert . Using the random-matrix representation  \mtam ,\tmam \ 
for the bare vertices, the   $L$-dependent part  \tver\ of the dressed  
 vertices can be calculated by means of a random matrix model
with a special potential. Namely, the generating function of the dressed
vertices
\eqn\gfdv{\CU [J]= \sum _{H=0}^{\infty} \sum _{n=o}^{\infty} 
{N^{2-2H-n} \over n!} \int 
 W^{(H)}(L_{1},...,L_{n}) \prod _{k=1}^{n} J(L_{k}){dL_{k} \over L_{k} }}
is equal to the free energy of the effective matrix model
\eqn\atwo{e^{\CU [J]}=\int d\Phi e^{(N{\rm tr} U(\Phi )
+\int _{0}^{\infty} {\rm tr } \Phi (L) J(L) dL/L}}
 with non-polinomial potential 
\eqn\aone{ U(\Phi )= U_{\c}(\Phi) + 
\beta \int _{0}^{\infty} {dL \over L} 
e^{-2P_{0}L} W^{(0)} (L){\rm tr} e^{L\Phi}}
The potential $U(P)$  is analytic in the the complex plane cut along 
the line $2P_{0}-P_{R} <P < 2P_{0}-P_{L} $; it can be represented 
as an infinite series in $\Phi $ with
radius of convergence $2P_{0}-P_{R}$. Its derivative is related to the planar
loop amplitude by
 \eqn\aonebis{ U'(\Phi)= \beta \hat W^{(0)} (2P_{0}-\Phi )  +U '_{\c}(\Phi)}

Geometrically,  the free energy $\CU [J] $ of the
effective matrix model   is the partition function of the gas of 
loops on a surface with fluctuating geometry with the
condition that  all {\it noncontractible}  loops are suppressed.
 For example, loops going around a handle or a boundary are forbidden.

The one-loop amplitude in the matrix model \atwo\  coincides
 in the planar limit $N \to \infty $ with
the string background \cleqm\ (the
  partition function of the loop gas on a disc)
\eqn\largn{\langle N^{-1} {\rm tr} e^{L\Phi} \rangle =  W^{(0)}(L)
,\ \ \ N \to \infty }
In order to prove this, it will be sufficient to show
 that \largn\ satisfies the loop equation \ccg . Indeed,
the planar loop amplitude $\hat W(P)$ in any matrix model is related to the
potential $U(P)$ by
\eqn\wpu{\eqalign{
{\rm Re}W(P)=  &-{1 \over 2} U'(P), \ \ P_{L}<P<P_{R} \cr
{\rm Im}W(P)= & 0, \ \ \ \ \ \ \ P<P_{L}\  {\rm or} \  P>P_{R}\cr}}
With the choice \aonebis\ of the potential, \wpu\ is equivalent to \ffr\ 
and therefore to \ccg  .

We are interested in the continuum limit of the dressed vertices
\eqn\cldv{w^{(H)}(\l_{1},...\l_{n})= a^{-(g+1)(n-2+2H)} \ W^{(H)}
(\l_{1}/a,...,\l_{n}/a), \ \ \ \ \ a \to 0}
Below we give a method for their evaluation using the random matrix
representation \atwo . 
The idea is to replace the potential \aone\ 
 with a simpler one  such that the corresponding matrix model
will generate  directly the  scaling limit \cldv\ of the dressed vertices.

Such a potential  will represent a function $u(z,M)$ with the following scaling
property
\eqn\scle{u(z,M)=\rho ^{\alpha} u(\rho z, \rho M)}

  Comparing \wpu\ and
the functional equations \scalir\  for the renormalized 
classical loop field $\hat w(z)$ , we see that a possible choice of the
scaling potential \scle\ is
\eqn\pottt{u(z)=\beta  \int _{0}^{\infty} {d\l \over \l} w(\l ) e^{z\l};
\ \ \ \  {d \over dz} u(z)= \beta \hat w(-z)}
Let us mention that
the inverse problem of finding the potential given the loop amplitude
has many solutions.
For example, the multicritical points of pure gravity can be obtained from
the polynomial potentials of Kazakov \kazm\ as well as from the non-polynomial 
singular potentials of Gross and Migdal \grmg . 


 We have shown that the  dressed vertices in the scaling limit \cldv\ can 
be calculated as the multi-loop correlators in the effective  matrix
model with potential \pottt 
\eqn\emmv{ \sum _{H=0}^{\infty} w^{(H)}(\l_{1},...,\l_{n})N^{2-n-2H}
=\int d\Phi e^{\beta  N \int _{0}^{\infty} (d\l / \l ) w(\l ) {\rm tr }
\exp (\l \Phi )  }
\prod _{k=1}^{n} {\rm tr }e^{\l_{k}\Phi}}
The one- and two-loop correlators in the planar limit are
\eqn\olp{w^{(0)}(\l) =  {1 \over \l } M^{g} K_{g} (M\l )}
\eqn\tlp{w^{(0)}(\l, \l ')= {\sqrt {ll'} \over l+l'} e^{-M(l+l')}}
The general formula for the planar multi-loop correlators in the matrix
model   \ajm  \mss  reads
\eqn\ddbbb{\eqalign {
w^{(0)}
(\l_{1},...,\l_{n})\ = &{ \sqrt{\l_{1}...\l_{n}}  
\over (\l_{1}+... \l_{n})}\ 
\Big(  - {\partial \over  \partial  
y} \Big)^{n-2} e^{-f(y; M ) (\l_{1}+...+\l_{n})}\cr
= & \sqrt{\l_{1} ... \l_{n} }
{\p^{n-2} \over \p y ^{n-2}} \int _{f(y; M )}
^{\infty}dz e^{-z(\l_{1}+...+\l_{n})}}}
where the derivatives are taken at the point $y=0$ and the function $f(y; M)$
is determined  by the diffusion equation (Appendix A) corresponding to the
potential \pottt 
\eqn\di{\eqalign{
y &= \lim _{N \to \infty} 
 \beta \int _{0}^{\infty} d\l \ w(\l ) \langle y | e^{z(-f(t)+(d/dt)^{2})}|
y \rangle \cr 
&= \int _{0} ^{\infty}{ d\l \over \l} \Big[\beta M^{g} K_{g}(M\l ) \Big]
\Big[ { e^{-f(y; M)\l} \over \sqrt {\l } } \Big]}}
We assume that the infinite constant produced by the singularity at small
distances $ (\l \sim a )$  is subtracted so that 
\eqn\nstm{f(0; M)=M}

The finite part of the integral can be extracted by means of analytic
continuation w.r. to a regulating parameter $\alpha $.
First we multiply the integrand by $\l ^{\alpha +1/2}$, then calculate
the integral \di\  in the domain ${\rm Re} \alpha >g$ where it converges, 
using the formula
 \ref\grrz{I.Gradshtein and I. Ryzhik, Table of
Integrals, Series, and Products, Academic press, 1965}      
\eqn\ddg{\eqalign{
&\int _{0}^{\infty} d\l \ \l^{\alpha - 1} e^{-z\l} K_{g}(M \l) \cr
&= \sqrt {{\pi \over 2M}} (M +z)^{1/2 -\alpha}
{\Gamma (\alpha +g) \Gamma (\alpha -g) \over \Gamma (\alpha +1/2)}\cr
 &  _{2}F_{1}(1/2+g,1/2-g;1/2+\alpha ; {M -z \over 2M}) \cr
& = \sqrt {{\pi \over 2M}} (M+z)^{1/2-\alpha} \sum_ {k=0}^{\infty}
{\Gamma (1/2+g+k) \Gamma (1/2 -g+k) \over \Gamma (1/2+\alpha +k) k!}
\Bigl ( {M - z \over 2M } \Bigr )^{k}   \cr}}
and finally perform analytic continuation to $\alpha = 1/2$.

 The result is
\eqn\conctd{\eqalign{
 y =& \beta M^{g}\sqrt {{\pi \over 2M}} (M+f)
 \sum _{k=1}^{\infty} {\Gamma (1/2 +g +k)
\Gamma (1/2 -g+k) \over k!(k-1)!} \Big( {M-f \over 2 M} \Big)^{k} \cr
=& 2(2\pi )^{3/2} (g^{2}-{1 \over 4}) M^{g+1/2} \cr
& \sum _{k=1} ^{\infty}
{1 \over k! (k-1)!}
{ \Gamma (k-1/2 +g) \Gamma (k-1/2 -g) \over \Gamma (1/2 +g)\Gamma (1/2 -g)}
\Big({M-f \over 2M}\Big)^{k} \cr }}
(the $k=0$ term disappears because of the $\Gamma (0)$ in the denominator.)

The parameter $y$ is the coupling constant of the matrix model;
in terms of the loop gas it is  the coupling of a special local operator,
to be discussed below.
 
\subsec{Some critical exponents}
  Until this moment we have considered two of the critical exponents of the
 loop gas model - the string susceptibility $\gst$ and the fractal
dimension of the loops $ D_{B}=1/\nu$ which
 is the same as the fractal dimension of
the boundary of the world sheet.  Now we are going to introduce a third
exponent $D_{C}$ which has the meaning of fractal dimension of the 
connected domains  bounded by loops. 

By construction, the vacuum energy of the matrix model
 in the 
planar limit   is the partition function of the gas of nonintersecting 
loops on the sphere, with one labeled domain.One  can imagine it as the
partition function of the gas of small spots (baby universes) floating
in the labeled domain.
A spot of perimeter $\l$ represents the partition function
 of the loop gas on a disc and have a Boltzmann weight $u(\l ) = \beta
w(\l )$. 
The size of these baby universes is of the order of the cut-off $a$
because of their entropy being divergent at $\l \to 0$.
The measure in the space of world-sheet metrics is tuned to the $m$-critical
phase of nonembedded random surfaces.  The parameter $y$ makes sense of
``cosmological constant'' for the labeled domain.

 Let $A_{C}$ and $A$ be the area  of the labeled connected domain and 
the total area of the world sheet, correspondingly. (Of course, the labeled
domain is equivalent to any other connected domain on the world sheet.) 
Then the fractal dimension of a single connected domain is
 determined by the way these two   areas
grow near the critical point $M=0$
\eqn\frdm{A_{C} \sim A^{D_{C}/2}}
The area of a single connected domain is measured by the matrix-model
puncture operator
\eqn\pun{\CP_{C}=-{\p \over \p y}}
The coupling constant $y$ of the matrix model scales as $M^{g+1/2}$; this
is  sufficient to determine the fractal dimension $D_{C}$.
It is convenient to consider the dense and the dilute phases separately.

\leftline{{\it (i)  Dense phase }}

In this phase 
\eqn\ppup{\CP_{C}= -{\p \over \p y} \Rightarrow  A_{C}, \
 \ \CP={\p \over \p \L} \Rightarrow A}
The cosmological constant $\L$ is related to $y$ by
$y=M^{g+1/2}  =
\L ^{1/2 +1/(4g)}$ and the fractal dimension of the connected area is
\eqn\frdi{D_{C}=1 +{1 \over 2g}}
For positive $\beta \ (g>1/2)$ all Boltzmann weights are 
also positive and the gas of loops  allows a statistical interpretation.
Then the area of a connected domain  diverges more slowly than the total area
\eqn\areas{A_{{\rm  conn}} \sim A^{1/2 +1/4g}}

In the dense phase $A$ coincides with the total length of the 
loops on the world sheet, and $A_{{\rm  conn}}
$ - with the total length of the loops
forming the boundary (hull) of a connected domain.

The scaling \areas\ is confirmed by the following argument based on the Coulomb gas picture.

The susceptibility $\p ^{2} \CU  / \p y^{2}$ of the matrix model can be 
interpreted as the two-point function of a special local operator defined on
the whole world sheet and not only on the labeled connected domain.
The two-point function of this operator 
 is zero when the two
points are in different domains and one if they are in the same domain.
This operator can be constructed as a vertex operator with electric charge
$e=g-1\pm 1/2$ \ref\dsds{B.Duplantier and H. Saleur, Phys. Rev. Lett. 
63 (1989) 2536; B. Duplantier, Phys. Rev. Lett. 64 (1990)493}.
 The corresponding gravitational scaling dimension
\eqn\locop{\delta _{0,1/2}={1/2-|g-1| \over 2g} = 1/2 -1/(4g)}
 is positive when $1/2<g<1$. The dimension  $1-\delta _{0,1/2} =D_{C}/2$
of the coupling constant of this
operator indeed coincides with the dimension of $y$.

\leftline{ {\it (ii) Dilute phase}}

In this phase the puncture operator $\CP _{C}$ of the matrix model
scales as $\L ^{-(g+1/2)/2}$. 
In the $m$-critical phase $\CP_{C}=A_{C}^{m/2}$ and the two 
areas  are related by
\eqn\nloopss{A_{C} \sim A^{D_{C}/2}; \ \ D_{C}=(2g+1)/m}
In the Coulomb gas picture, the constant $y$ is coupled to the same local 
operator with gravitational scaling dimension
\eqn\dimbul{\delta _{0,1/2}={1/2 -|g-1| \over 2}=3/4-g/2}
This time $1-\delta _{0,1/2}=(m/4)D_{C}$.

Knowing the area of a connected domain, we can evaluate the characteristic
number of connected domains
 on the world sheet. By the Euler theorem, this is also the number of loops
 {\cal N} (including the most external loop representing the boundary). 
Therefore
\eqn\nloops{\eqalign{
{\cal N} = A/A_{C}  & \sim  A^{1-D_{C} }\cr
&=
\cases{
A^{1/2 -1/(4g)},\ \  & dense phase \cr
A^{(m-g-1/2)/m},\ \ & $m$-critical dilute phase \cr}}}
In the interval $1/2 <g<3/2$ where the loop gas has statistical interpretation
(all Boltzmann weights are positive, including these related to the measure
over random surfaces) the total number of loops tends to infinity when
the critical point $M=0$ is approached. The density of loops is proportional
to $1/A_{C}$ and goes down to zero at the critical point.

The number of loops becomes one at the half-integer points
$g=m-1/2, m=1,2,3,...$ which are the solutions of the equation $\beta =0$.
These are the multicritical points of pure gravity \kazm \ where
the only loop is the boundary of the world sheet.
 At $m=1$  and $m=2$ the dimension $\delta _{0,1/2} $ of the local operator
coupled to $y$ is zero and therefore $\CP_{C}=\CP$. In the higher 
multicritical points with $\beta =0$  the dimension of the operator $P_{C}$
fits the 
 dimension  $\delta _{m-1,1}$
 of the Kac table of the model $(2,2m-1)$
\eqn\dida{\delta _{0,1/2}={1/2 -|m-3/2| \over 2}= {(m-1)-(m-1/2)-|m-3/2|
\over 2} =\delta _{m-1,1}}
Therefore $P_{C}$ can be identified with the operator with the most negative
dimension in the $m$-critical regime of pure gravity \stau \bdks \cgm .

\subsec{Feynman rules}
Now we are able to formulate 
the Feynman diagram technique
for the $ADE$ strings in a more  explicit way.  

The vertices  \pvr \ and \vert\ read, in the continuum limit
\eqn\ccve{v^{(H)}(x|\l_{1},...,\l_{n})
=\big(\kappa S_{x} \big) ^{2H-2+n}
w^{(H)}(\l_{1},...,\l_{n})}
\eqn\cvce{v^{(H)}(p_{1},...,p_{n}|\l _{1},...\l_{n})=
\kappa ^{n-2+2H}N^{(H)}_{p_{1}...p_{n}}w^{(H)}(\l_{1},...,\l_{n})}
where $w^{(H)}$ are the loop amplitudes in the one-matrix
model with potential $u(z)=\beta \hat w(-z) $, $\hat w(z)$ given by \ffx ,
and the renormalized string interaction constant $\kappa$ is defined by
\eqn\kap{\kappa = a^{-\nu(2-\gst)} \kappa _{0}=a^{-g-1}\kappa _{0}}

The explicit form of the vertices without handles  ($H=0$)
 is given by \conctd\ and \ddbbb . To our knowledge, the explicit
formula for the general macroscopic loop amplitudes in the one-matrix model 
is not yet found.  Below we give an argument allowing to imagine the 
general form of the matrix-model loop amplitudes with $H>0$. 

 In the general case the loop amplitudes
in the effective matrix model can be considered as loop amplitudes for
the gaussian model dressed by tadpoles representing the nongaussian part of
the potential  (see eq. \tver ).  From the explicit
 form of the gaussian loop amplitudes (Appendix B) we conclude that
\eqn\vertc{w ^{(H)}(\l_{1},...,\l_{n})=\sqrt {\l_{1}...\l_{n}}
\  P_{n}^{(H)}(\l_{1}+...\l_{n}) e^{-M(\l_{1}+...+\l_{n})}}
where $ P_{n}^{(H)}(\l)$ is a polynomial of degree $n-3+3H$
with coefficients depending on $g$ and $M$
\eqn\poly{P^{(H)}_{n}(\l)= \sum _{k=0}^{n-3+3H} A_{n;k}^{(H)}(g, M) \l^{k}}
(Note that $n-3+3H$ is the complex dimension of the moduli space for
surfaces with $H$ handles and $n$ punctures).

Let us denote the $n$-loop amplitude in the continuum limit by
\eqn\loopfield{\langle \prod _{k=1}^{n} \Psi  _{x_{k}}(\l _{k}) \rangle }
where  $\Psi _{x}(\l)$ is the loop  operator creating a boundary of length $\l$
at the point $x$. For  the tree approximation $(H=0)$  we  will use the symbol 
$\langle ... \rangle _{0}$. 
The v.e.v. of the loop operator is, in the tree approximation,
\eqn\tdplc{\langle \Psi _{x} (\l) \rangle = {1 \over \l}M^{g} K_{g}(M\l) S_{x}}

The two-loop correlator satisfies the equation 
\eqn\eqntwlp{\langle \Psi _{x_{1}}(\l _{1}) \Psi _{x_{2}} (\l _{2}) \rangle
=\sum _{x} C_{x_{1}x}\int _{\l=0}^{\infty} d\l\  w^{(0)}(\l _{1},\l) \langle 
\Psi _{x} (\l)\Psi _{x_{2}}(\l_{2}) \rangle }
where $w^{(0)}(\l,\l')$ is the  two-loop amplitude in the effective matrix
 model. The latter  is 
diagonalized by Bessel functions \grrz
\eqn\diagct{w(\l,\l')={\sqrt {\l \l'}  \over \l +\l'}e^{-M(\l+\l')}
=\int _{0}^{\infty} dE \langle \l |E \rangle {1  \over 2\cosh \pi E}
\langle E|\l' \rangle }
\eqn\diagcti{\langle \l | E \rangle ={2 \over \pi}
 \sqrt {\pi E \ \sinh (\pi E)} K_{iE}(M\l); \ \ \ 
 K_{iE}(M\l) = \int _{0}^{\infty} d\tau e^{-M \l \cosh \tau} \cos (E\tau)}
where the states  $\langle \l | E \rangle $ form a complete orthonormal
system \mss . On the other hand, the connectivity matrix is diagonalized
by
\eqn\dddia{C_{xx'}=\sum _{p} V_{(p)}^{x} 2\cos (\pi p) V_{(p)}^{x'}}
where the sum includes the allowed momenta $p=m/h$. Thus the wave functions
of the closed string states diagonalizing the two-loop correlator
are labeled by the ``energy-momentum'' $(E,p)$
and have a factorized form
\eqn\eigsts{ \langle E,p|\l, x \rangle = V^{x}_{(p)} \sqrt {E\ \sinh (\pi E)}
K_{iE}(M\l)}

The equation for the two-loop correlator becomes algebraic after the
diagonalization and its solution  is \Imult
\eqn\tlcrdg{\langle \Psi _{x}(\l) \Psi _{x'}(\l') \rangle _{0} 
=\int _{0}^{\infty} dE  \sum _{p} \langle \l,x|E,p \rangle {1 \over
2\cosh (\pi E) -2\cos (\pi p)} \langle E,p|\l ',x' \rangle }
For $p=1/2$ (pure gravity) and $p=1/3$ (Ising model) the two-loop 
correlator has been  calculated  from  matrix models
\mss , \ref\mmmsss{G. Moore and N. Seiberg, ``From loops to fields in 2$D$
quantum gravity'' , preprint RU-91-29  and YCTP-P19-91}. 
One can check that  \tlcrdg  \  reproduces the 
expressions found for these particular cases.

Remarkably, the two-loop correlator is universal; it depends on the model 
only through its spectrum of allowed momenta. This makes possible to map the 
Hilbert spaces of the $ADE$ strings onto the Hilbert space of the SOS string
whose spectrum of excitations covers the whole interval $-1<p\le 1$.
The two-loop amplitude is essentially the propagator in the string diagram
technique. The latter is given by the continuum limit of \eignst
\eqn\egtc{G_{(p)}(\l, \l')= \int _{0}^{\infty } dE 
\langle \l| E \rangle \tilde G(E,p) \langle E|\l ' \rangle }
\eqn\eignstc{
\tilde G(E,p)={2\cosh (\pi E) \cos (\pi p) \over 
\cosh (\pi E) - \cos (\pi p)}}

The $(E,p)$ space has the geometry of a semi-infinite cylinder. It is
represented by the infinite $(E,p)$ plane factored by the relation
of equivalence
\eqn\facr{(E,p) \equiv (E,p+2)}
The periodicity of the propagator \eigsts\  is a direct consequence of the
discreteness of the coordinate space $X$.

It follows from the explicit form of the eigenstates \diagcti\  that
 the $\tau$-parametrization  of the boundary 
cosmological constant $z$ (eq. \parr ) 
provides a local coordinate dual to the 
energy $E$.

 The configuration space of the string field
theory  is a direct product of the $X$ and $\tau$ spaces.
The relation between the  $\l$ and $\tau$  representations has been
discussed in ref. \mmmsss . 
It has been argued that 
 the $\tau$ variable makes sense of the time variable in the Das-Jevicki 
approach, and that the
 transformation from $\l$ to $\tau$  is analogous to the
B\"acklund transformation in the Liouville theory.
This transformation  reads explicitely   \mmmsss
\eqn\grmoore{
\eqalign{
 K_{iE}(M\l) & = \int _{0}^{\infty} d\tau e^{-M \l \cosh \tau} \cos (E\tau) \cr
 {\cos  {E\tau} \over E \sinh {\pi E}} &= \int _{0}^{\infty} {d\l \over \l }
e^{-M \l \cosh {\tau}} K_{iE}(M\l) \cr }}
and does not relate delta-function normalized bases of wavefunctions.

 Let us consider the 
case when the target space is the discretized real line $\Z \subset \R$.
 The inverse 
propagator \eignstc\  acts  in the $ (x,\tau)$ space 
as a differential operator of infinite order
\eqn\prxt{ \hat G^{-1}=\ [ \cosh(\pi \p /\p x )]^{-1}-
[\cos (\p / \p \tau )]^{-1}}
This operator can be also interpreted as a finite-difference operator in
the $x$ and $i\tau$ directions. Thus Minkowski rotation $\tau \to 
t=i\tau $ of the
$\Z$ string is described by a lattice Hamiltonian.
The saddle-point equation \ham\  means that
  the classical string background is annihilated by the Hamiltonian \prxt .
 \smallskip
It  will be 
 convenient to apply the Feynman rules directly in the $(E,p)$ space.
The Fourier image of the $\l$-dependent part of the vertices \vertc\ 
can be found using the integration formula 
\eqn\erepv{\int _{0}^{\infty} {d\l \over \l} \sqrt {\l} e^{-M\l}
\l ^{k} \ K_{iE}(M\l) = {\sqrt {\pi } \over  \big( 2M \big)^{k+1/2}}
{ \Gamma (k+1/2+iE) \Gamma (k+1/2 -iE) \over \Gamma (k+1)}}
which follows from  \ddg .  The result is
\eqn\zbxz{\tilde v^{(H)}(p_{1},...,p_{n}|E_{1},...,E_{n})=
\kappa ^{n-2+2H} N_{p_{1}...p_{n}} \tilde w^{(H)}(E_{1},...,E_{n})}
\eqn\vespa{\eqalign{&
 \tilde w^{(H)}(E_{1},...,E_{n})  = 
\int _{0}^{\infty} \prod _{k=1}^{n}{d\l_{k} \over \l_{k}}
\langle E_{k}| \l_{k} \rangle w^{(H)}(\l_{1},...,\l_{n}) \cr
& =\sum _{m=0}^{n-3+3H}\sum _{m_{1}+...+m_{n}=m;\ m_{k}\ge 0}
m! \ A^{(H)}_{n;m}(g,M) \cr &\times
 \prod _{k=1}^{n} \sqrt {E_{k} \sinh (\pi E_{k})} \ 
{\Gamma (1/2 \ +m_{k}+iE) \Gamma (1/2 \ +m_{k}-iE) \over m_{k}!\ m_{k}! }\cr}}

The vertices in the $(E,p)$ space are given by the product of the r.h.s. of
\vespa\ and \vrp . Each vertex represents a symmetric polynomial in $E_{1},
...,E_{n}$ times a factor $\sqrt{\pi E_{k} \sinh (\pi E_{k})}/\cosh (\pi 
E_{k})$ for each line. 

It is convenient to absorbe the non-polynomial
factors in the propagators; then the new propagator and vertices have the
following form

$$\tilde G_{new} (E,p) = {\sqrt{ \sinh (\pi E) } \over
 \sqrt {\pi E}  \cosh (\pi E)} 
{ 2\cos (\pi p) \cosh (\pi E) \over \cosh (\pi E) -  \cos (\pi p)}
 {\sqrt{  \sinh (\pi E)/E } \over \cosh (\pi E)}$$
\eqn\newpr{= G_{\heartsuit}(E,p) -  G _{\heartsuit}(E,1/2) }
\eqn\bleve{G_{\heartsuit}(E,p)= { \sinh (\pi E) \over \pi E}
{ 2 \over \cosh (\pi E) - \cos (\pi p)} = \sum _{n=-\infty}^{\infty} 
{1 \over E^{2}+(p+2n)^{2}}}
$$\tilde V_{new}^{(H)}(p_{1},...,p_{n}|E_{1},...,E_{n})
=N^{(H)}_{p_{1}...p_{n}}\sum _{m=0}^{n-3+3H}
\sum _{m_{1}+...+m_{n}=m;\ m_{k}\ge 0}$$
\eqn\newvr{ m! \ A^{(H)}_{n;m}  \prod _{k=1}^{n} {E_{k} \over  ( m_{k}!)^{2}}
\prod_{s=0}^{m_{k}-1}[E_{k}^{2} +(s+1/2)^{2}]}
With the Feynman rules \newpr , \newvr\  the evaluation of the loop 
amplitudes can be done as in an ordinary quantum field theory \foot{ The
form of the propagator \newpr\ suggests an interpretation of the string
field as a collection of two modes: a bosonic relativistic particle 
with compactified momentum space in the space-like direction and
a ghost excitation (twist) with spectrum of momenta $p= \pm 1/2$}. 	
The Greens functions of this effective field theory are the correlation
functions of the Fourier-transformed loop field 
\eqn\furpsi{\tilde \Psi (E,p) =\sum _{x} (S_{x})^{2}\int _{0}^{\infty}{d\l
 \over \l}  \chi _{(p)}^{x}
\Psi _{x}(\l)}
\eqn\furpsii{\tilde \Psi (E,p)_{new}= {  \sqrt{E} \cosh (\pi E)
 \over \sqrt{ \sinh (\pi E)}
}\tilde \Psi (E,p)}

Let us calculate, for example, the three-loop correlator
 of the loop field $\Psi_{(p)}(\l)=
\sum _{x} \chi _{(p)}^{x} \Psi _{x}(\l)$. We start with the corresponding
Greens function in the $(E,p)$ space which is given by a single Feynman diagram
\eqn\eaea{\langle \prod _{k=1}^{3} \Psi (E_{k},p_{k}) \rangle  _{0} = 
C_{p_{1}p_{2}p_{3}} M^{-(1+g)} \prod _{k=1}^{3}\tilde G(E_{k},p_{k})}
Then we integrate w.r. to the $E$ variables associated with the three legs
according to the formula
\eqn\leggs{\int _{0}^{\infty} dE {E \sinh (\pi E) \over \cosh (\pi E)-
\cos (\pi p)} K_{iE}(M\l)= M\l K_{1-|p|}(M\l)}  
to find
\eqn\thrlp{\langle \prod_{k=1}^{3}\Psi _{(p_{k})}(\l_{k}) \rangle 
=\L^{-(1-\gst /2)}
C_{p_{1}p_{2}p_{3}}\prod _{k=1}^{3} \l_{k} K_{1-|p_{k}|}\big(M\l_{k}\big)}
In the particular case $p_{1}=p_{2}=p_{3}=p_{0}$ corresponding to three loops
associated with the identity operator, eq. \thrlp \ is in accordance with the
the formula conjectured in \Inonr .  

\subsec{Spectroscopy}

The spectrum of on-shell states is given by the poles of $\tilde G(E,p)$.
They form the light cone in the $(E,p)$ cylinder (Fig. 8)

\eqn\lghtspc{E= \pm i\varepsilon _{n} (p),\ \ \  \varepsilon _{n} (p)
 = |p+2n|; \ n=0,\pm 1, \pm2 ,...}

The periodicity in the momentum space is a consequence of the discreteness 
of the coordinate $X$ \ space. If the $X$-space is compact, then the 
momentum space is discrete as well and the on-shell
states form an infinite two-dimensional lattice on the space-time
cylinder.

     \epsfxsize=100pt
    \vskip 20pt
    \hskip 100pt
    \epsfbox{ 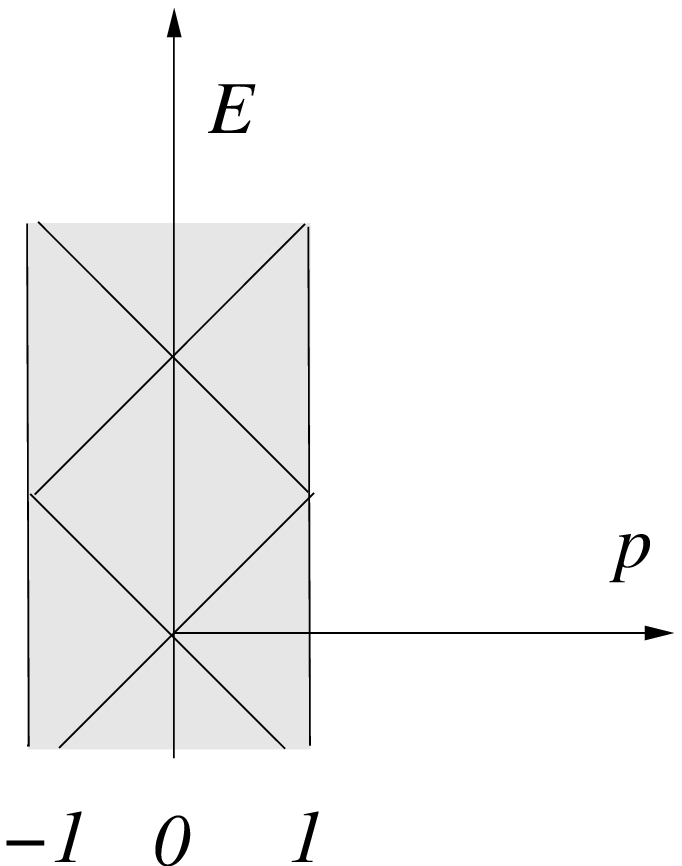   }
    \vskip 5pt
    
    \centerline{\ninepoint Fig. 8:    The light cone for a compact 
    momentum space. } 
    \centerline{The edges of the strip
correspond to identical momenta $p=\pm 1$.}
     
     \vskip  10pt

     Each on-shell state creates a ``microscopic loop'' on the world sheet 
and corresponds to a local scaling operator. The sequence of poles 
corresponding to a given momentum $p$ defines the set of local operators
$\CO_{(p);n};n=0,\pm 1,\pm 2,...$
having nonvanishing correlations with the vertex operator
\eqn\vropa{V_{(p)} \ \equiv \CO _{(p);0} = \sum _{x} \chi ^{x}_{(p)}\CP _{x}}
where $\CP_{x}$ is the  operator creating a puncture in the world sheet 
with coordinate $x$.  
These operators are the coefficients in the expansion of the off-shell 
loop field $\Psi _{(p)}(\l)$ as a series of on-shell wave functions  \mss 
\eqn\expnsion{\Psi _{(p)}(\l)= \sum _{n=-\infty}^{\infty} 
(|p|+2n) M^{||p|+2n|}I_{||p|+2n|} (M\l) \CO _{(p);n}} 

The expansion of the 
two-loop correlator as a sum of products of on-shell wave functions reads
\eqn\expsi{\langle \Psi _{(p)}(\l)\Psi _{(p)}(\l')\rangle =
 \sum _{n=-\infty}
^{\infty}{|p|+2n \over \sin (\pi p)}
K_{|p|+2n}(M\l)\ I_{|p+2n|}(M\l') , \ \ \ \l'<\l}
Eqs. \expnsion\ and \expsi\ imply
\eqn\xespn{\langle \CO_{(p);n} \Psi_{(p')}(\l)\rangle = M^{||p|+2n|}
K_{|p|+2n}(M\l)}
A direct derivation of (Laplace transform of ) \xespn\  using
 the loop equations can be found in \Iade\ and \Inonr .
 
The gravitational dimension $\delta _{n}(p)$  of the local operator
$\CV_{(p);n}$ can be found from \xespn 
\eqn\dmsns{M^{||p|+2n|}=\L ^{\delta _{n}(p)- \gst /2} \Rightarrow 
\delta _{n}(p)={||p|+2n|-|p_{0}| \over 2\nu}}
where $p_{0}=g-1$ is the background momentum \psol \  (see  \crex\ , \crexm ).

The  dimensions of the operators  $\CO_{(p);n}$
 with $p=kp_{0}$ , $k$ - integer,
are contained in the KPZ-DDK spectrum \grdm
\eqn\kpzddk{ \delta (\CO_{rp_{0};n})= \delta _{k+2n, k}}

\subsec{Partition function of the noninteracting string}
The Feynman rules in the continuum limit are generated by the functional
integral
\eqn\ffin{\eqalign{
 e^{F[J]}  = & \int d \Psi e^{\CA [\Psi ]} \cr
 \CA[\Psi ]  = & \cr
  &  -{1 \over 2} \sum _{p} \int _{0}^{\infty} {d\l 
\over \l} \Psi _{(p) }(\l) \int _{0}^{\infty} dE \langle \l |E \rangle
\Big( {1 \over  \cos\ \pi p} -{1 \over \cosh\ \pi E} \Big) 
\langle  E| \l ' \rangle \Psi _{(p)}(\l ')\cr
 \ \  & + \sum _{n=2}^{\infty} \sum _{H=0}^{\infty } \sum _{p_{1},...,p_{n}} 
 {1 \over n!}\int _{0}^{\infty} \kappa ^{n-2+2H} N^{(H)}_{p_{1},...,p_{n}} \cr 
& \ \  P^{(H)}_{n} (\l_{1}+...+\l_{n})
 \prod _{k=1}^{n}
\Psi _{(p_{k})} (\l_{k}) {d\l_{k} \over \sqrt {\l_{k}}} \cr }}
which formally defines the string field theory with target space $X$.
The partition function $ F^{(1)}$ of the free string (topology of a torus) is
 closely related to  the vacuum energy of this  field theory.
 
The loop expansion for the embedded surfaces with the topology of a torus
involves loop configurations with contractible and noncontractible loops.
It is easy to see that the  logarithm of the  determinant 
  of the kernel $G_{xx'}(\l,\l ')$  is equal to twice  the contribution
of the surfaces in the loop expansion containing at least one noncontractible
loop. Indeed, if we return to the original diagram technique (subsec. 2.5) and
retain only the gaussian part of the action, the contribution of the
gaussian fluctuations of the $\Psi$-field to the vacuum energy will be given
by the sum of all  Feynman diagrams with the topology of a tree with one 
cycle. Each cycle corresponds to a closed  path in $X$ and consists of
even number of points (bare vertices).
 A cycle with $2k$ points should be taken with a symmetry factor $2k$.

Let us denote by $F_{[k;p]}$ the contribution of the toroidal surfaces
($H=1$) containing exactly $k$ noncontractible loops with momentum $p$.
 Then the  partition  
function  for fixed momentum $p$ is equal to
\eqn\pfpf{\eqalign{
F_{(p)} =&  \sum _{k=0}^{\infty} F_{[k;p]  } \cr
=& F_{[0]} +  {1 \over 2} \sum _{k=1} ^{\infty} \ {1 \over 2k} 
 \sum _{p}\int_{a}^{\infty} {d\l \over \l} \int _{0}^{\infty} {dE \over
\pi }  \langle l|E \rangle
\Big( {\cos (\pi p) \over \cosh (\pi E)} \Big) ^{2k} \langle E|\l \rangle \cr
 =& F_{[0]} - \int _{a} ^{\infty} {d\l \over \l}
|\langle \l |E \rangle |^{2}  \log \Big( 1 \ - \ 
{\cos ^{2} (\pi p) \over \cosh ^{2} (\pi E)} \Big) \cr }}

The contribution of the surfaces without noncontractible loops is equal to the 
trace of the identity operator in the $X$-space (= the number of its points)
 times the $1/N^2$-correction
to the free energy of the effective matrix model considered in the beginning
of this section. A simple calculation (see, for example Appendix C of \grmg )
leads to
\eqn\fser{F_{[0]}= {g-1/2 \over 24}\log (aM)}

The  integral over the length $\l$ in  \pfpf\ yields a factor $\log [1/(aM)]$
 (this is the diagonal 
value of the kernel of the regularized identity operator in the 
$E$-space) and the $E$-integration gives
\eqn\prttor{
 \int _{0}^{\infty}
dE  \log \Big( 1 - {\cos ^{2} (\pi p) \over
 \cosh ^{2} (\pi E)}\Big) ={\pi \over 2}  (1/2-|p|)^{2} }
Inserting this in \pfpf\ we find for contribution to the partition function 
of an excitation with momentum $p$
\eqn\ppff{ F_{(p)} =
 \log (aM){    
     g-1/2 + 6(1/2 -|p|)^{2}\over 24} }

Let us give some examples.
\smallskip
\leftline { {\it (i) The SOS string compactified on $\Z_{2h}$}}

Summing over the allowed momenta
$p=m/h;\  m=0,\pm 1,...,\pm(h-1),h$ in 
the embedding space $X=\Z _{2h}$  we find
\eqn\znn{
F_{2h}(g)=\log (aM)
 \Big( {h^{2}+2 \over 24h } + 2h \  {g-1/2 \over 24} \Big)
= \log (aM) \Big(
{gh+1/h \over 12} \Big)}

The $\Z_{2h}$ string is well defined if the background momentum $p_{0}$
belongs to the spectrum of allowed momenta. Therefore the possible values
of $g$ are
\eqn\lats{g=m\pm {k \over h}; \ \ \ m={\rm integer}, 0 \le k <h}
The partition function of the string embedded in $\Z_{2h}$ is symmetric 
under the duality transformation 
\eqn\duall{g \to 1/g,\ h \to 1/h , \ M \to M^{g}}
relating the dense and dilute phases. Due to the different scaling in these 
two phases, the cosmological constant is invariant under the duality
transformation.  The self-dual point is $g=1, h=1$. It corresponds to
the level-one representation of an $SU_{2}$ current algebra.
The Kosterlitz-Thouless transition may occur at $ g=1,h=2$ and its dual
point  $g=1,h=1/2$. The case $h=2$ describes the $\Z_{4}$ model which 
is a particular case of the  Ashkin-Teller model. At this point the
magnetic (vortex) operator with discontinuity $m=4$ (= the period in the
$X$ space) becomes marginal: its conformal dimension \dmop\ is 1. 
The dual model is formally the $\Z_{1}$ model having as a target space  
a graph with one point and one loop. The space of paths in this space is not 
trivial because each step is made by traversing the loop in one of the two 
possible directions. This $\Z_{1}$ string is equivalent to the $n=2$ limit
of the $O(n)$  string introduced in \Ion . The vortex operators in the
$O(n)$ string are the electric operators in the $\Z_{1}$ string.
The spectrum of momenta consists of all integers and the marginal
operator is the vertex operator with $p=2$ (= the period in the momentum
space).

\smallskip
\leftline{
{\it (ii) The RSOS strings}}

The embedding space $X=A_{h-1} $ has spectrum of momenta $p=m/h;\  
m=1,2,...,h-1$ and the partition function reads \foot{Here we restrict
ourselves to the unitary theories with $0<g<2$.}
\eqn\brty{\eqalign{
F_{A_{h-1}}= &
{(h-1)(h-2) \over 12h} \log (aM) \cr
 = & {h-2 \over 24}\log (a^{2(1-1/h)}\L)
,\ \ \  {\rm dense \ phase} \  (g=1-1/h) \cr }} 
\eqn\brtyy{ \eqalign{
F_{A_{h-1}}= & {h(h-1) \over 12 h} \log (aM)\cr  
 = & {h-1 \over 24} \log (a^{2} \L),
\ \ \ {\rm  dilute\  phase} \  (g=1+1/h) \cr} }
In both critical regimes the partition function of the $A_{h-1}$ string
is equal to the difference between the partition functions of the $\Z_{2h}$
and $\Z_{2}$ strings
\eqn\cgrps{ F_{A_{h-1}}=  {1 \over 2} \Big( F_{2h}(g=1\pm 1/h)
-F_{2}(g=1\pm 1/h) \Big) }
The relation \cgrps\ takes place even before performing the sum
 over the world sheet geometries \Iade . 
\smallskip

In a similar way one can calculate the  partition functions of the $D$ and $E$
strings.  In the $A$ and $D$ cases the expressions will coincide with these
obtained from the matrix models in the formulation of M. Douglas 
\mike\  by Di Francesco and Kutasov \ref\dik{P. di Francesco and
D. Kutasov, Nucl. Phys. B 342 (1990) 589}. 
All these partition functions can be expressed as linear
combinations of partition functions of $\Z_{n}$ strings \Iade .
 The formulas are
the same as these for the regular lattice \fsz . 
 The Coulomb gas calculation of 
the genus-one partition function of the $ADE$ strings
was presented in \bk .

\newsec{Concluding remarks}

We have shown that the propagation and interactions of strings embedded in
a discrete one-dimensional target space are described by string field
theory with propagator \eignstc\ and vertices \vespa .  We were not able
to calculate explicitely the vertices with higher topology but they are 
defined unambiguously by the matrix integral \atwo . 

Our diagram technique based on the loop-gas representation
of the IRF models,  provides a natural decomposition of the moduly space into
elementary cells. The vertices represent the
 topology-changing amplitudes  involving no
propagation in the target space. The fact that the vertices are 
themselves loop amplitudes guarantees the ``stringy'' large order behaviour
\ref\stiv{S. Shenker, Rutgers preprint RU-90-47 (1990)}.  
 The vertices included in a 
Feynman diagram cut a number of holes in the moduli space such that the
rest of it is a direct product of one-dimensional spaces (one for each
propagator). The degenerated surfaces appear when the proper time
for  some of the propagators becomes infinite
 (the proper time is measured by the number of
the noncontractible loops along the corresponding cylindric surfaces).
This decomposition of the moduli space is well defined only if all proper
times are nonzero (i.e., if there is at least one noncontractible loop
in each of the channels. The situation when one of the proper times
vanish is taken into account  by replacing the two involved vertices
with a new vertex of higher topology. 
 Thus the interaction of the closed string is described by a collection
infinitely many vertices, one for each topology. This is possibly a
general feature of the perturbative expansion for any string theory
\ref\zw{H. Sonoda and B. Zweibach, Nucl. Phys. B 331 (1990) 592;
B. Zweibach, MIT preprint CPT-1926, December 1990}. 
\smallskip
It would be interesting to compare our diagram technique for the
loop amplitudes  to the one
for the string embedded in $\R$. 
The two-loop correlator for the string
embedded in $\R$  \mss   \ref\Mmre{G. Moore, Rutgers preprint 
RU-91-12}, \mmmsss . 
\eqn\mooore{\tilde G(E,p)={E \over \sinh (\pi E)}
{1 \over E^{2}+p^{2}}}
is compatible with \tlcrdg\ only in the limit of small $p \   \ and\  \  \ E$.
The propagator of the $\R$-string contains an infinite sequence of poles
at integer imaginary values of $E$ which have been interpreted as the 
special stationary states \foot{ Since these  ``states''  can be
 elliminated by redefinition of the vertices, their physical meaning is
not very clear to us.}  discovered by A. Polyakov \ref\plst{A. Polyakov,
Mod. Phys. Lett. A6 (1991) 635}. These states are created by infinitesimal 
motion in the $X$ space; in the Coulomb gas picture they are constructed by
taking the derivatives of the $x$-field. Therefore, such states should
not exist in a string theory with discrete target space.
The special states in the $\R$-string are related to the closed classical
trajectories in the imaginary time direction.

The infinite spectrum of states in our case comes from the fact that
 the discreteness of the $X$-space leads to periodicity in the
momentum space. As a consequence, the light cone maps an infinite set 
of energies to the same momentum. These energies make the tower
of gravitational descendents of    the vertex operator with given momentum.
The classical motion in the imaginary time direction is gouverned by a 
finite-difference Hamiltonian.

The fact that the difference between
the embedding spaces  $\Z$ and $\R$ survives in the continuum limit can be
explained. 
 If we construct the D=1 string as an infinite
 chain of coupled matrices \ref\grkl{
D. Gross and I. Klebanov, Nucl. Phys. B 344 (1990) 475} \ref\pris{G.
 Parisi, Phys. Lett. B 238 (1990) 213}, then the $\R$-string will describe the
 low temperature phase of the matrix chain and the
$\Z$-string will describe the Kosterlitz-Thouless point where the distance
between two nearest neighbours on the chain remains finite in the continuum 
limit. The change in the critical behaviour of the matrix chain
 near the Kosterlitz-Thouless
point is due to the liberation of the angular excitations \ref\grkle{
D.Gross and I. Klebanov, Nucl. Phys. B 354 (1991)459}  
\ref\voldim{
V. Kazakov and D. Boulatov, ENS preprint, LDT-ENS-91/24
and KUNS-1094 HE (TH) 91/14, August 1990}

\centerline{{\bf Acknowledgements}}
I thank M. Douglas, P. Di Francesco, B. Duplantier, G. Moore, A. Polyakov,
S. Shenker,
M. Staudacher and A. Zamolodchikov for many stimulating discussions.
I am grateful to J.-B. Zuber for a critical reading of the manuscript.

\appendix{A}{The method of orthogonal polynomials}

The integral over Hermitean $N \times N$ matrices 
\eqn\appone{e^{N^{2}F}=\int d\Phi e^{N {\rm tr}U(\Phi)}}
can be written as an integral w.r.t. the eigenvalues $\phi _{n},n=1,...,N,$
of the matrix $\Phi$
\eqn\apptwo{e^{N^{2}F} = \int \prod _{k=1}^{N} d\phi _{k}e^{U(\phi _{k})}
\prod _{i <k} (\phi _{i}- \phi _{k})^{2}}
This integral can be interpreted as the scalar product of the wave function 
of $N$ fermions with itself. The one-fermion wave functions are of the form
\eqn\ortp{\langle \phi|n \rangle =  P_{n}(\phi) e^{U(\phi)/2}}
where $P_{n}(\phi ) $ is a polynomial 
of degree $n$. The polynomials are fixed by the condition 
of orthonormality
\eqn\ortn{\langle m|n \rangle =\int d\phi P_{m}
(\phi )P_{n}(\phi) e^{  U(\phi )} = \delta _{mn}}
The ground state of the system of fermions is equal to the antisymmetrized
product of the states $|n\rangle , n=1,...,N$
\eqn\groundst{\Xi(\phi_{1},...,\phi_{N})= {1 \over N!} \sum_{\sigma \in S_{N}}
(-1)^{[\sigma]}\langle 1|\phi _{\sigma _{1}} \rangle ...
\langle N| \phi _{\sigma _{N}} \rangle }

 The  operator $\Phi$ is represented in the  space of one-fermion states
is represented by the
tridiagonal Jacobi matrix which we denote by the same letter
\eqn\jacob{\Phi |n\rangle = R_{n+1}|n+1 \rangle + R_{n} |n-1 \rangle
+ S_{n} |n \rangle }
The nonzero matrix elements $S_{n}$ and $R_{n}=A_{n-1}/A_{n}$ of this
matrix are fixed by the recurrence relations
\eqn\rec{\langle n|\Phi  U'(\Phi)|n\rangle = (2n+1)/N}
\eqn\recbis{\langle n |  U'(\Phi) |n \rangle =0 }
which can be interpreted in terms of diffusion of a particle with $n$-dependent
hopping parameter.
In the limit $N \to \infty$ we can replace $R_{n},S_{n}$ by
\eqn\limit{R=\lim _{n \to \infty} R_{n}, S= \lim _{n \to \infty} S_{n}}
and the resolvent of the Jacobi matrix is given by the propagator of a 
one-dimensional relativistic particle
\eqn\resolv{\eqalign{
\CR _{nn'}(P)& = \langle n|{1 \over P-\Phi}|n' \rangle \cr
&= \Bigg({P-S - \sqrt {(P-S)^{2}-4R^{2}} \over 2R}\Bigg)^{|n-n'|} 
{1  \over \sqrt {(P-S)^{2}-4R^{2}}} \cr}}

Another useful quantity is the inverse Laplace transform of \resolv\
\eqn\loopi{ \langle n|e^{L\Phi}|n' \rangle = e^{SL} I_{|n-n'|}(2RL)}
where $I_{n}(z)$ is a modified  Bessel function.
%
%
 All quantities characterizing 
the radial part of the random matrix can be expressed in terms of \resolv\ or
\loopi . For example, the one- and two-loop correlators are given by 
\eqn\wilsloop{W(L)=  {1 \over N} \sum _{n=1}^{N}
\langle n| e^{L\Phi}|n \rangle = e^{SL}{1 \over N} \sum _{n=1}^{N}
I_{0}(2R_{n}L)}
\eqn\twoloopi{ W(L,L')= \sum _{n>N,n'<N} \langle n|e^{L\Phi}|n' \rangle
 \langle n'|e^{L'\Phi} |n \rangle }

In the limit $N \to \infty$  we can introduce the continuous variable
\eqn\xiks{y=1-{n \over N}}
and consider $R_{n}$ and $S_{n}$ as  functions   of  $ y $:
 \eqn\huy{ R_{n}=R(y), S_{n}=S(y); \ R(0)=R, S(0)=S}
  These functions
are determined
by \rec\ ,\recbis\ . Using the explicit expression \resolv\ for the resolvent
we find
\eqn\rnsn{
\int _{P_{L}}^{P_{R}}dP {PU'(P) \over \sqrt{(P-S)^{2}-4R^{2}}} = \pi(1- y) }

\eqn\rnsnbis{
\int _{P_{L}}^{P_{R}}dP {U'(P) \over \sqrt{(P-S)^{2}-4R^{2}}} = 0 }
The positions of the branchpoints of $W$ are therefore given by
\eqn\branch{P_{L}=S-2R,\ \ P_{R}=S+2R}
 It follows by \wilsloop\  that
\eqn\rx{ W(L)  = -\int _{0}^{1} dy \  e^{S(y)L} I_{0}(2R(y)L)}

\eqn\rxx {W(P)   = - \int _{0}^{1} dy \  {1 \over \sqrt{(P-P_{L}(y))
(P-P_{R}(y))}}}

The two-loop correlator is obtained immediately from \twoloopi\ 

\eqn\propl{ W(L,L')=e^{S(L+L')} \sum _{n=1}^{\infty} n 
I_{n}(2RL) \  I_{n}(2RL')} 
Its Laplace image  depends only on the positions of the
endpoints of the cut and is given by
\eqn\cormat{W(P,P')={\sqrt{(P-P_R)(P'-P_L)\over(P'-P_R)(P-P_L)}+
\sqrt{(P'-P_R)(P-P_L)\over(P-P_R)(P'-P_L)} - 2
  \over 4(P-P')^{2}}.
 }

In the continuum limit 
\eqn\kur{ L=\l /a, \ \ \ P=P_{R}^{*} +az, \ \ \ P_{R}(y) =P_{R}^{*} -af(y), \ \ \ f(0)=M }
eqs. \rnsn\ and \rnsnbis\  imply
\eqn\sclt{\int _{f(y)}^{\infty} {dz \over \sqrt{a(z+M)}} u'(-z) = y }
where $u(z)=U(P_{R}^{*}+az)$.  
If the potential $u(z)$ scales as $a^{\alpha}$,
the dependence on the cutoff $a$ can be  elliminated by
rescaling $  u(z) \to a^{\alpha} u(z), y \to a^{\alpha -1/2}y$.
Eq. \sclt\ is then the Laplace image of \di\ for the potential \pottt\ . 

Finally, the two-loop correlator in the scaling limit reads
\eqn\contl{w(\l , \l ') = \sqrt{\l} \ {e^{-M(\l +\l ')} \over \l + \l '}
\sqrt{\l '}}
\eqn\contp{w(z,z')= {{\sqrt{z+M} \over \sqrt{z'+M}} + 
{\sqrt{z'+M} \over \sqrt{z+M}} -2   \over 4 \ (z-z')^{2}} =
-{\p \over \p z } {\p \over \p z'}\log (\sqrt{z+M} + \sqrt{z'+M})}

%

%

In the case of gaussian potential $U(\Phi )=- \Phi ^{2} /2$ the one-loop
amplitude can be calculated explicitely in all orders in $\kappa $ in the
continuum limit \ref\fdunp{F. David, unpublished} \ref\brkz{E. Br\'ezin and
V. Kazakov, Phys. Lett. B 236 (1990) 144}
%
\eqn\aci{
w(\l) = \int _{-\infty}^{-M} dy \langle y| e^{l(y- \kappa ^2(\p /\p y)^{2})}|
y \rangle  
=\l ^{-3/2} e^{-M\l + \kappa \l ^{3}}.}

%

\listrefs
\bye